\documentclass{aa}  

\usepackage{txfonts}
\usepackage{graphicx,lipsum,overpic}
\usepackage{subfigure}
\usepackage{amssymb,amsmath}

\newcommand{\ignore}[1]{}

\def\aap {{A\&A}}

\def\apjl {{ApJL}}

\def\apjs {{ApJS}}

\def\mdot{\dot{M}}
\begin{document}

   \title{Vertical gas accretion impacts the carbon-to-oxygen ratio of gas giant atmospheres}

	\titlerunning{ Vertical gas accretion and C/O }

   \author{Alex J. Cridland\inst{1}\thanks{cridland@strw.leidenuniv.nl}, Arthur D. Bosman\inst{1}, \& Ewine F. van Dishoeck\inst{1,2}
          }

	\authorrunning{ A.J. Cridland, A.D. Bosman, \& E.F. van Dishoeck }

   \institute{
 $^1$Leiden Observatory, Leiden University, 2300 RA Leiden, the Netherlands \\ $^2$Max-Planck-Institut f\"ur Extraterrestrishe Physik, Gie{\ss}enbachstrasse 1, 85748 Garching, Germany
}

   \date{Received \today}


  \abstract
  {
Recent theoretical, numerical, and observational work have suggested that when a growing planet opens a gap in its disk the flow of gas into the gap is dominated by gas falling vertically from  a height of at least one gas scale height. Our primary objective is to include, for the first time, the chemical impact that accreting gas above the midplane will have on the resulting C/O. We compute the accretion of gas onto planetary cores beginning at different disk radii and track the chemical composition of the gas and small icy grains to predict the resulting carbon-to-oxygen ratio (C/O) in their atmospheres. In our model, all of the planets which began their evolution inward of 60 AU open a gap in the gas disk, and hence are chemically affected by the vertically accreting gas. Two important conclusions follow from this vertical flow: (1) more oxygen rich icy dust grains become available for accretion onto the planetary atmosphere. (2) The chemical composition of the gas dominates the final C/O of planets in the inner ($<$ 20 AU) part of the disk. This implies that with the launch of the \textit{James Webb Space Telescope} we can trace the disk material that sets the chemical composition of exoplanetary atmospheres.
}

   \keywords{astrochemistry, planets and satellites:atmospheres, planets and satellites: gaseous planets, planets and satellites: formation, protoplanetary disks, planets and satellites: individual: Jupiter
               }

    \maketitle
%

\section{Introduction}

The link between the measured carbon-to-oxygen ratio (C/O) in the atmosphere of a planet to its formation history has evolved significantly since it was proposed in \cite{Oberg11}. We have seen this connection is complicated by both the physics of planet formation, and the chemical and physical evolution in protoplanetary disks.

On the planet formation side of the problem, planets will migrate through their natal disk via Type-I migration before the planet opens a gap \citep{GoldTrem79,W91,Ward1997} and after through Type-II migration \citep{LP86}. These migrating planets will sample a wide range of chemical compositions if planet growth is slow. Meanwhile the C/O of the gas and ice change drastically; through changes in the gas temperature and density (see for ex. \citealt{Crid16a}), long time-scale gas-grain chemical reactions \citep{SW11,Reboussin2015,Eistrup2018,Schwarz2018}, and the dynamical evolution of the gas and dust \citep{Furuya2014,Yoneda2016,Bosman2017b,Krijt2016,Krijt2018,Booth2019}. 

Here we investigate another possible complication on the interpretation of the chemical characterization of exoplanetary atmospheres. One generally limits oneself to the chemistry along the disk midplane when discussing the chemical properties of the material that is accreted into gas giant atmospheres. However recent theoretical \citep{Morbidelli2014,Batygin2018}, numerical \citep{Szul2014}, and observational \citep{Teague2019} evidence has shown that gas accretion into the gap caused by a growing planet is predominantly vertical. The planet feeds on gas that falls from heights between one and three scale heights \citep{Teague2019}, rather than along the disk midplane.

\begin{figure}
\includegraphics[width=\columnwidth]{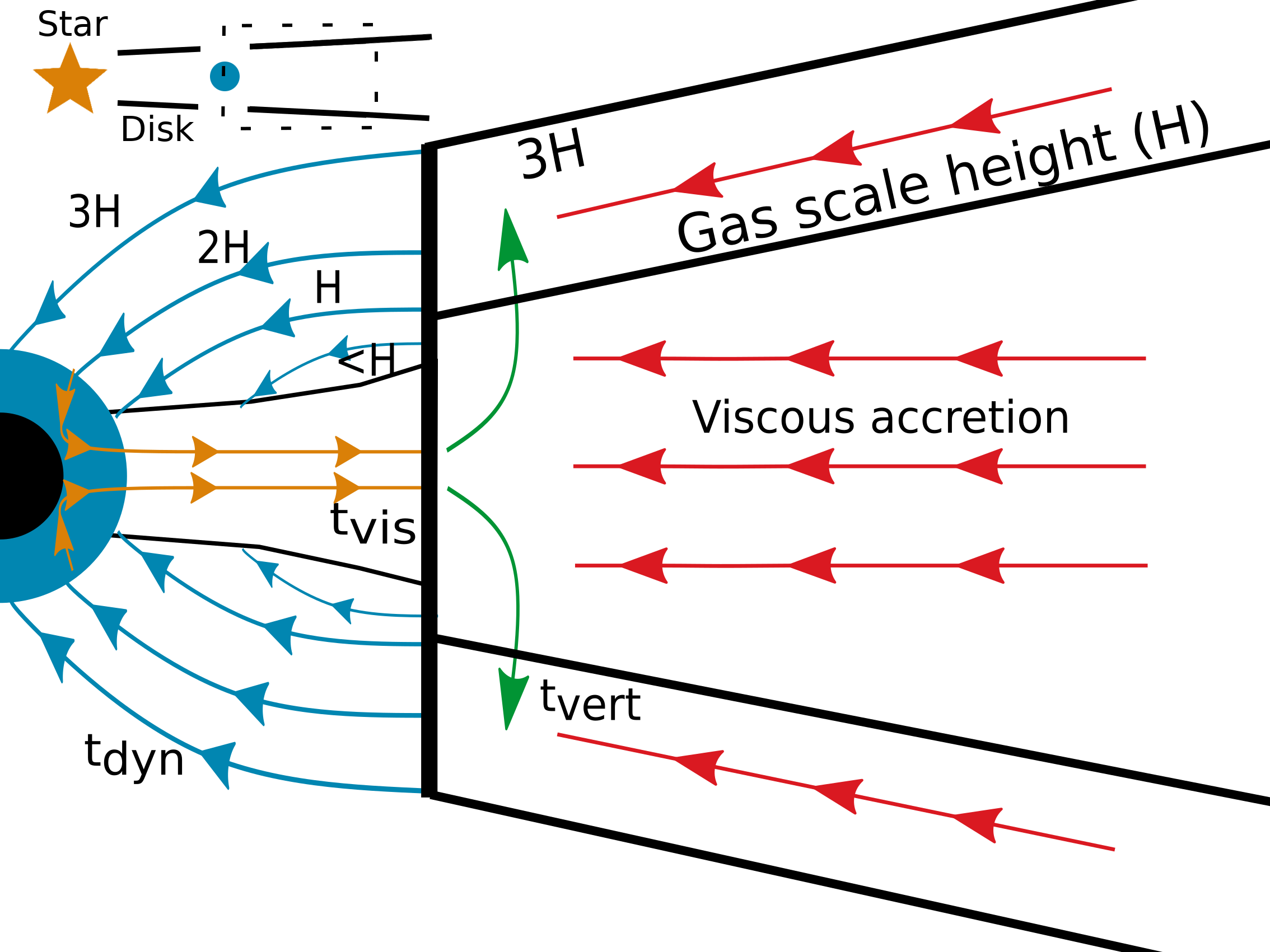}
\caption{Here we show schematically the physical system that drives the vertical accretion assumed for this work. When a planet is large enough, it opens a gap in its natal protoplanetary disk (top left corner). When this occurs, \citet{Batygin2018} argues that the majority of gas accretion onto a planet falls vertically from between one and three scale heights (blue arrows). Some of this gas is incorporated into the growing planet, while some recycles back into the surrounding protoplanetary disk through the circumplanetary disk (yellow arrows). Gas falling into the gap from below one scale height also feeds the circumplanetary disk and does not reach the planet. Once in the surrounding protoplanetary disk there is vertical flow (green arrows) which reestablishes hydrostatic equilibrium near the edge of the gap. Meanwhile the gap is also supplied by new gas accreting radially from radii outside the current location of the planet (red arrows). The relevant timescales of these processes are labelled and discussed later in this work.}
\label{fig:intro01}
\end{figure}

This flow was described by \cite{Morbidelli2014} as `meridonial circulation', and is shown schematically in Figure \ref{fig:intro01}. Meridonial circulation begins with gas flowing over the traditional edge of the planetary gap due to the fact that at a given radius the planetary gravity is weaker high above the midplane. This gas loses its hydrostatic balance and free falls towards the disk midplane and the growing planet (blue arrows in Figure \ref{fig:intro01}). Some of this gas is accreted onto the planet, while the rest flows back into the surrounding protoplanetary disk through a circumplanetary disk (yellow arrows). The final step in meridional circulation is to re-balance the gas hydrostatic balance in the protoplanetary disk near the gap edge, pushing gas back up to heights where it can flow into the gap again (green arrows).

The question then becomes: is the gas between one and three scale heights chemically different from the gas on the midplane? And if so, do these differences alter the resulting C/O in the atmospheres of giant planets? In this work we compute the chemical structure of the disk volatiles using the two-dimensional thermo-chemical code DALI \citep{Bruderer2012,Bruderer2013}, coupled to a full gas-grain chemical network to compute the chemical composition and temperature of the gas and the dust.

We assume that the chemical composition of the proto-planetary atmosphere is either dictated strictly by the composition of midplane material, or is allowed to be determined by the chemical structure of gas above the midplane after the gap has been opened. To distinguish between these two scenarios we name the former model `1D accretion' (that is midplane only), while the latter is called the 2D accretion model.

In what follows we describe our chemical and gas accretion models in section \ref{sec:chemmethods} and \ref{sec:plntmethods} respectively. We report the resulting atmospheric C/O for a set of planets with a wide range of initial disk locations in section \ref{sec:results}. We discuss in section \ref{sec:disc01} their importance and conclude in section \ref{sec:conclusion}.

\section{ Two-dimensional chemical structure }\label{sec:chemmethods}

\subsection{Methods}

Our disk model is based on a typical protosolar system, which describes the pre-main sequence phase of low-mass stars (up to $\sim$ M$_\odot$). Such a system would be a typical pre-main sequence phase for the bulk of stars around which the majority of known exoplanets are found (see for example \citealt{Kep14}). Meanwhile the bulk of directly imaged planets have been observed around stars of at least 1.5 solar masses \citep{Stone2018,Nielsen2019}. These solar systems are best described by Herbig Ae pre-main sequence stars, which typically have higher disk masses, gas temperature, and UV flux. Broadly speaking we do not expect the fact that gas accretes vertically to depend on the particular disk model, however the chemical impact of this flow could change depending on the disk model.

The DALI \citep{Bruderer2012, Bruderer2013} code is used to calculated the physical and chemical structure of the disk, with a disk mass of 0.025 $M_\odot$ around a 1 $M_\odot$ star. The stellar spectrum is composed of two black bodies with temperatures of 4250 and 10000 K and luminosities of 1 and 0.15 $L_\odot$, representing the proto-star and the accreting column of gas respectively. The disk surface density is distributed according to a tapered power-law,  which is a self-similar solution to the viscous equation \citep{LB74}. The vertical gas density profile is a Gaussian with a disk radius ($r$) dependent scale-height:
\begin{equation}
\frac{H(r)}{r} = h(r) = 0.1 \left(r/35 ~\mathrm{AU}\right)^{0.3}.
\end{equation}
The dust is split into two populations (small grains, 0.005 - 1 $\mu$m and large grains, 0.005 -- 1 mm) and is distributed following \cite{Trapman2017}. The small grains (1\% of the solid mass) are assumed to be fully mixed with the gas, and follow the same vertical distribution, while the large grains (99\% of the solid mass) are settled with a scale height that is $h_\mathrm{large} = 0.2 h_\mathrm{gas}$. 

The gas and dust structure is calculated using DALI. This structure is then post processed with the full gas-grain chemical network from \citet{Bosman2018CO}. This network is based on the gas-phase \textsc{Rate12} network \citep{McElroy2013}\footnote{\url{http://www.udfa.net}} and grain surface reactions from the OSU network\footnote{http://faculty.virginia.edu/ericherb/research.html} \citep{Garrod2008gas-grain}. The wavelength dependent UV photo-dissociation and ionisation reactions, based on the cross-sections of \cite{Heays2017}, that are used in DALI have been added to the network. This chemical network is evolved for 1 Myr using the standard comic ray ionization rate of $\zeta_{CR} = 10^{-17}$ s$^{-1}$. In the grain surface chemistry, the quantum tunneling barrier width is set to 1 \AA, and the diffusion to binding energy ratio for ice species is assumed to be $0.3$.

We assume that the chemical composition of the refractories (dust, planetesimals) do not contribute to the atmospheric composition of our synthetic planets. Hence we do not compute any chemical evolution for the refractory component, such as C-burning \citep{Anderson2017}. We compute the impact of accreting refractories in \cite{Crid19b} and found that they have a strong impact on the final C/O of giant planet atmospheres. In this work we are investigating the impact of vertical accretion, and hence we do not wish to overshadow this effect with a more model-dependent process like refractory solid accretion. As explained below, however, we do include the ice component frozen onto the micron-sized dust grains, since these grains are dynamically linked to the gas flow and we expect that they are accreted along with the gas.

The focus of this paper is to investigate the chemical effects of 2D vs. 1D accretion. We use a disk model that is static in both physical and chemical structure to highlight the effects of the different assumptions on the accretion physics. The complex thermo-chemical model ascertains that we include all the important physical and chemical gradients that are expected in the disk, even though the locations of these gradients would change with time.


\begin{table}
\centering
\caption{Parameters for the DALI model}
\begin{tabular}{l c}
\hline\\
\multicolumn{2}{c}{Stellar parameters}\\
\hline\\
Stellar mass & 1 $M_\odot$\\ 
Stellar luminosity & 1 $L_\odot$\\ 
Stellar temperature & 4250 K\\ 
Accretion luminosity & 0.15 $L_\odot$\\ 
Accretion temperature & 10000 K\\
X-ray luminosity & $10^{30}$ erg s$^{-1}$ \\  
\hline\\
\multicolumn{2}{c}{Disk parameters}\\
\hline\\
Disk mass & 0.025 $M_\odot$ \\
Critical radius ($R_c$) & 35 AU \\
Surface density slope ($\gamma$) & 1\\
Scale-height at $R_c$ ($h_c$) & 0.1 rad\\
Flaring angle ($\psi$) & 0.3 \\
Small dust size & 0.005--1 $\mu$m \\
Small dust fraction & 0.01\\
Large dust size & 0.005--1 mm \\
Large dust fraction & 0.99\\
Large dust settling factor & 0.2 \\
Cosmic-ray ionisation rate & 10$^{-17}$ s$^{-1}$ \\
C/H & $1.35\times 10^{-4}$ \\
O/H & $2.88\times 10^{-4}$ \\
N/H & $2.14\times 10^{-5}$ \\
\hline
\end{tabular}
\end{table}

\subsection{Results}

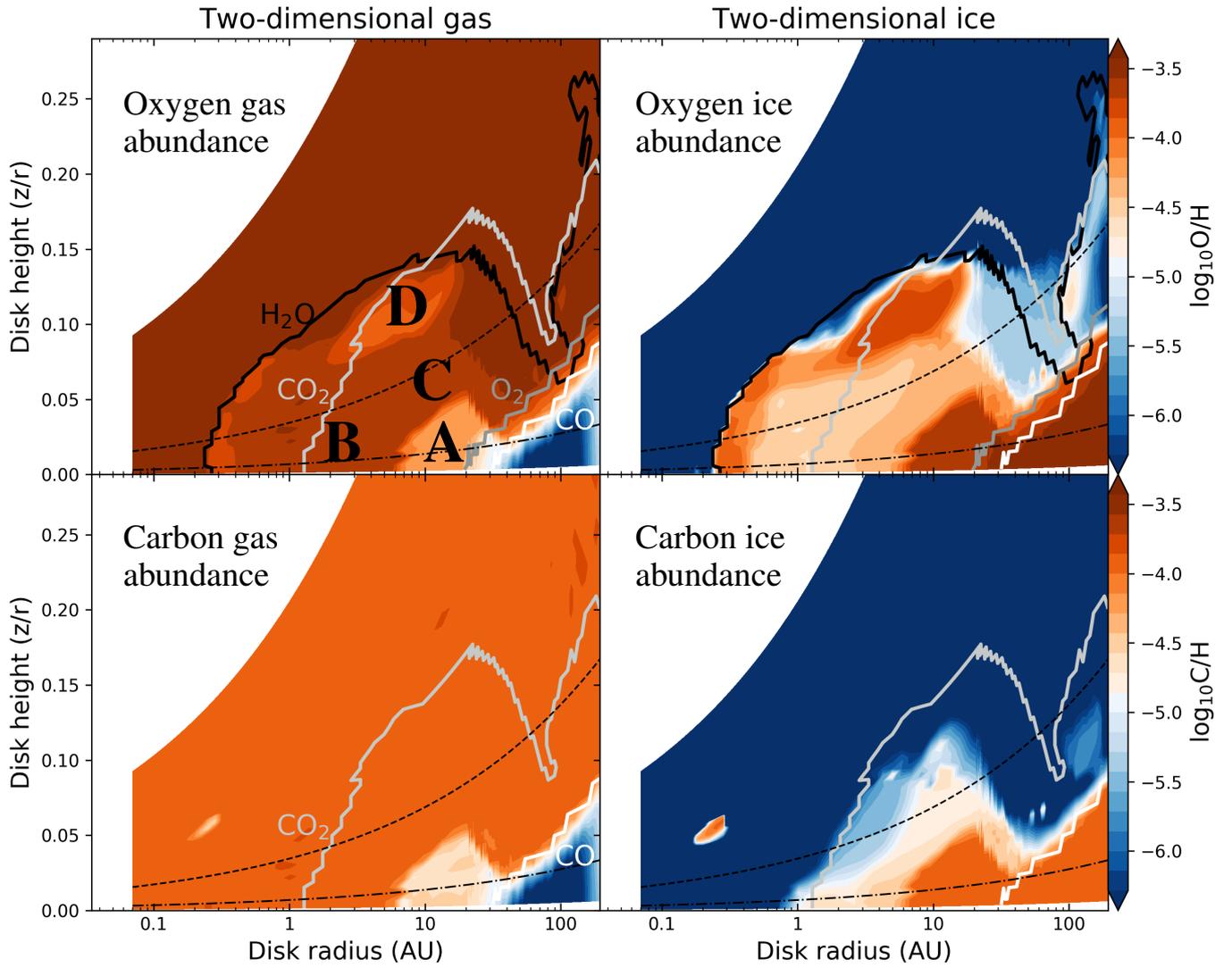
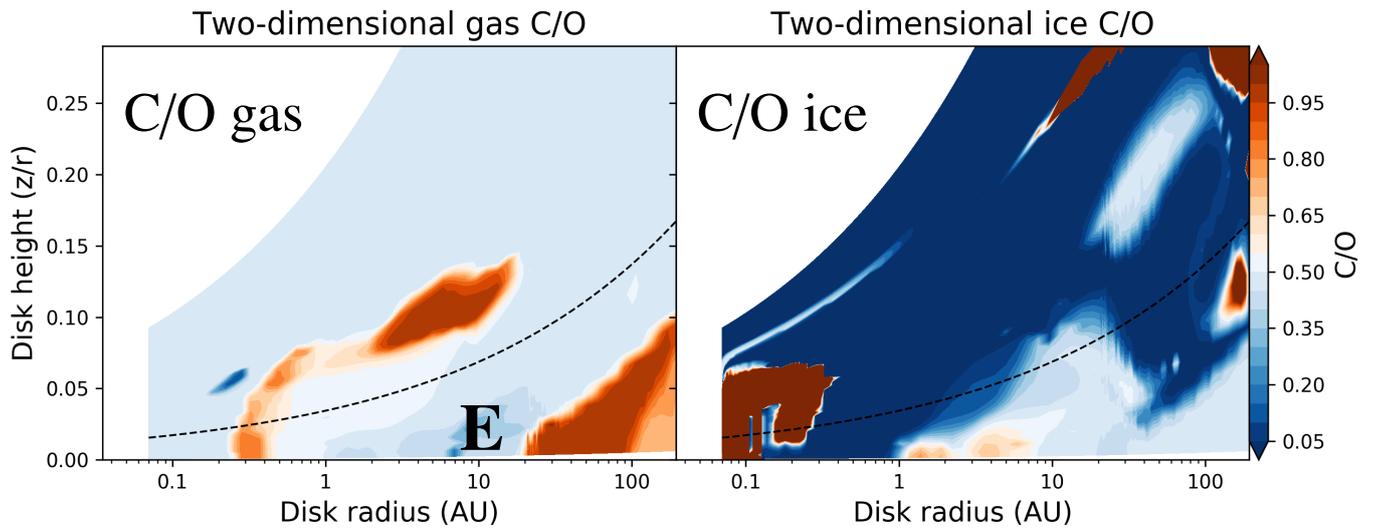
\begin{figure*}
\centering
\subfigure[ Total oxygen and carbon partition between the gas and ice. We note the gas scale height with a thin dashed line, while we show the scale height of the large grains with a dot dashed line. We show the water, carbon dioxide, carbon monoixde, and molecular oxygen ice lines with the black, silver, white, and grey contours respectively. To the right and below these curves, the disk is sufficiently cold that these molecular species freeze out onto the dust grains. The letter labels relate to locations of chemical interest (see text). ] {
\begin{overpic}[width=\textwidth]{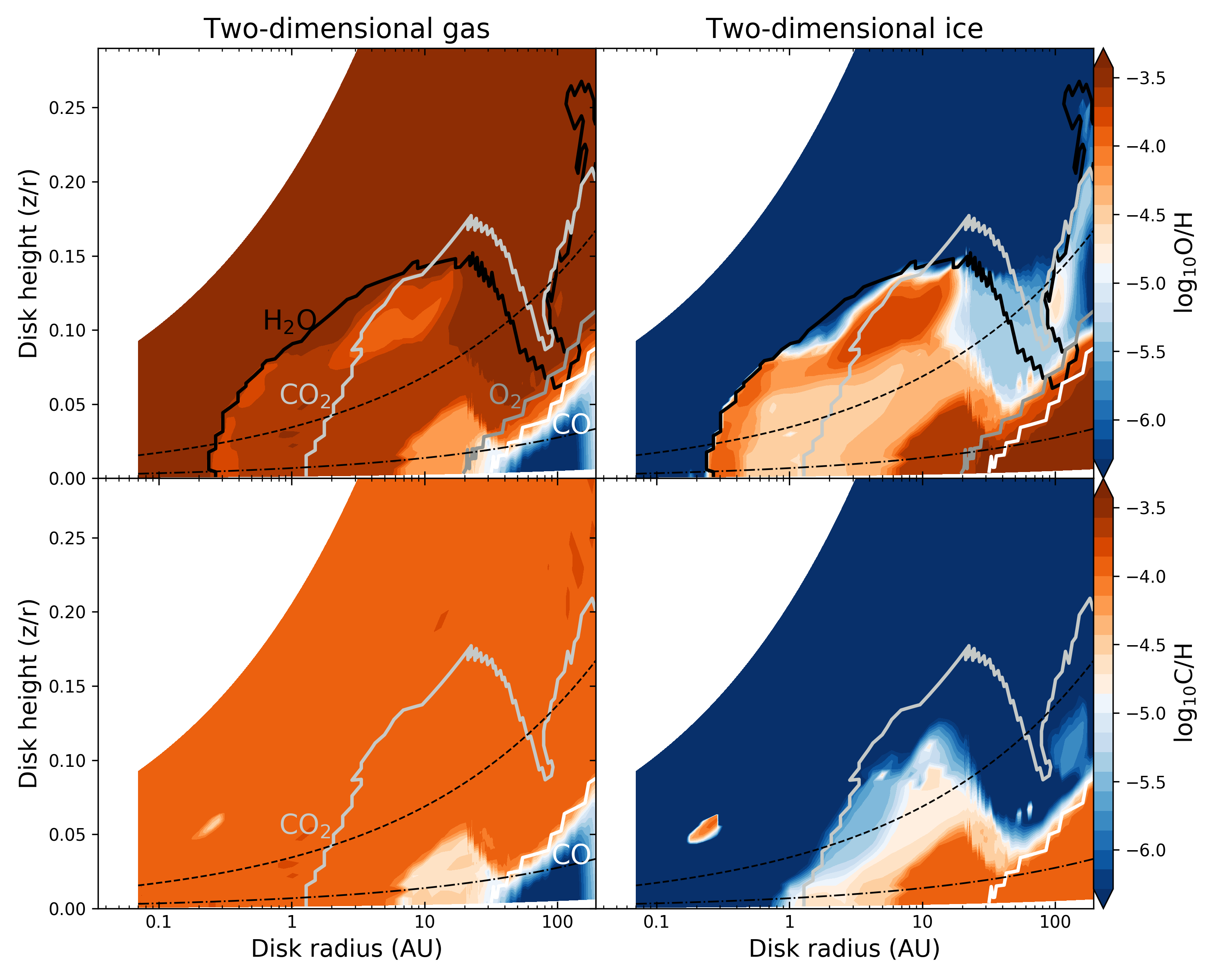}
\put(10,35){ \Large Carbon gas }
\put(10,32){ \Large abundance }
\put(10,70){ \Large Oxygen gas }
\put(10,67){ \Large abundance }
\put(51,35){ \Large Carbon ice }
\put(51,32){ \Large abundance }
\put(51,70){ \Large Oxygen ice }
\put(51,67){ \Large abundance }
\put(34,42){ \Huge {\bf A} }
\put(26,42){ \Huge {\bf B} }
\put(33,47){ \Huge {\bf C} }
\put(31,53){ \Huge {\bf D} }
\end{overpic}
\label{fig:chemmodel01}
}
\subfigure[ Carbon-to-oxygen ratio in the gas and ice. The white contour is solar C/O, while blue and orange represents sub- and super-solar C/O respectively. ]{
\centering
\begin{overpic}[width=\textwidth]{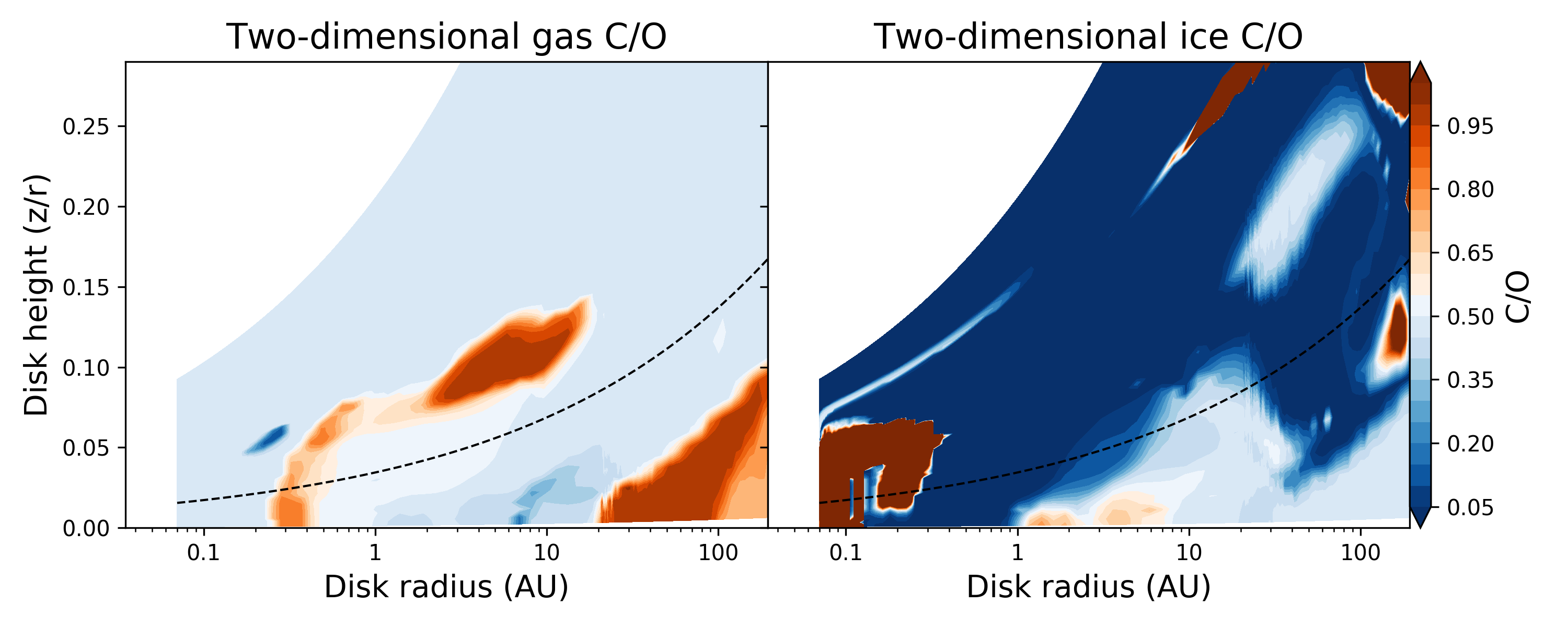}
\put(9,30){ \huge C/O gas }
\put(50,30){ \huge C/O ice }
\put(33,7){ \Huge {\bf E} }
\end{overpic}
\label{fig:chemmodel02}
}
\caption{Chemical model of one quadrant of our adopted disk produced by DALI.}
\end{figure*}

In Figure \ref{fig:chemmodel01} we show the distribution and partition of carbon (bottom row) and oxygen (top row) between the gas (left column) and ice (right column). The majority of features are related to the volatile ice lines, and hence we show the ice surfaces for the primary carbon and oxygen carriers in the disk. To the right and below the grey-scale contours the dust temperature is sufficiently low that the corresponding volatile has frozen out onto the dust grains. At a few scale heights above the midplane, photo-processes alter the abundance of both the gas phase (via photodissociation) and ice phase (via photodesorption) of volatiles. As such the colour contours describing the partition tend to deviate from our simple estimate of the ice surface location based on drawing a contour between regions in the disk where the absolute abundance of a molecule $X$ in the gas phase $X_{\rm gas} < X_{\rm ice}$ and where the opposite is true. 

The bulk chemical changes in the disk are related to the ice surfaces of water (black curve in figure \ref{fig:chemmodel01}) and the CO ice surface (white curve in figure \ref{fig:chemmodel01}). Between these surfaces a significant amount of chemistry can occur, which locally changes the partition of carbon and oxygen between the gas and ice \citep{Oberg11,Eistrup2016}. At temperatures above, but within $\sim 10$ K of the freezing temperature of CO, gaseous CO will briefly be absorbed onto the dust grain and react with an OH radical before it returns to the gas phase. This reaction produces CO$_2$ on the grain, which stays frozen at these temperatures (see Figure \ref{fig:app} or \citealt{Eistrup2016}). Overall this leads to a reduction in the oxygen abundance in the gas phase (see A. in Figure \ref{fig:chemmodel01}) and the gaseous carbon, while locally increasing the oxygen and carbon abundance in the gas phase.

At higher gas and dust temperatures, at radii nearer to the inner edge of the CO$_2$ ice surface (see B. in Figure \ref{fig:chemmodel01}), the OH radicals are more mobile on the dust grain. This means that the radical is more likely to find atomic or molecular hydrogen to produce water ice, reducing the efficiency of the reactions discussed above. Similarly above the midplane (see C. in Figure \ref{fig:chemmodel01}) the absolute abundance of CO gas drops with the bulk density. However the absolute abundance of atomic hydrogen remains mostly constant because it is produced by the dissociation of molecular hydrogen by cosmic rays \citep{Bosman2018CO}, which again reduces the efficiency of CO$_2$ production.

At heights farther above the gas scale height (see D. in Figure \ref{fig:chemmodel01}) we find a further increase in the oxygen abundance of the ice due to the destruction of CO gas by reactions with ionized helium. This leads to the production of water ice, and gaseous methane (see Figure \ref{fig:app} and also \citealt{Furuya2014,Bosman2018CO}).

The impact of these chemical reactions can be seen in the C/O of both the gas and ice in Figure \ref{fig:chemmodel02}. Where CO can be converted to CO$_2$ ice, the C/O in the gas is reduced relative to the surrounding gas (see E. in Figure \ref{fig:chemmodel02}), while the C/O in the ice increases. The partitioning of carbon and oxygen is important since the relative accretion rates of solids and gas can differ significantly (see for ex. Cridland et al. submitted).

In our disk, volatile C/O never exceeds unity apart from regions of the disk where C/H and O/H in the ice gets infinitesimally small. This occurs inside of 0.3 AU below H/r $\sim$ 0.08, and in the outer (r $>$ 10 AU) disk where H/r $\sim$ 0.3. These regions of the disk are characterized by high temperatures and high ionization. Other than in these regions, C/O is not larger than unity because carbon never exceeds oxygen in abundance in neither the gas nor ice.

\section{ Methods: Gas accretion and chemical inheritance }\label{sec:plntmethods}

\subsection{ Planetary growth model }

For this work we skip the initial core growth as we assume that its construction has little impact on the chemical composition of the atmosphere. Instead, we begin the growth of our synthetic planets at a core mass of 10 $M_\oplus$, which represents a typical mass for the beginning of runaway gas accretion \citep{Ikoma2000,IL08}, and compute the rate of gas accretion through a set of models described below.

When the planet is still embedded in the surrounding protoplanetary disk then its accretion is limited by the rate at which the gas can lose its gravitational energy. Under this limit, the gas accretion rate is limited to the Kelvin Helmholtz timescale:\begin{align}
t_{\rm KH} &= 10^c {\rm yr} \left(\frac{M_{\rm plnt}}{M_\oplus}\right)^{-d},
\label{eq:gas01}
\end{align}
where the constants $c=7$ and $d=2$ were determined by comparing a population synthesis model of planet formation with the population of known exoplanets \citep{APC16a}. The growth of the protoplanet is thus: $dM_{\rm plnt}/dt = M_{\rm plnt}/t_{\rm KH}$.

At a certain point this timescale leads to an unstable accretion of available gas, potentially adding more mass to the planet than is gravitationally bound to the planet. The gas that is gravitationally bound to the planet is enclosed within the Bondi radius, which describes the ratio between gravitational and thermal energy:\begin{align}
R_{\rm B} = \frac{GM_{\rm plnt}}{c_s^2},
\label{eq:gas02}
\end{align}
where $G$ is the gravitational constant, and $c_s = \sqrt{k_B T/\mu m_H}$ is the gas sound speed. The rate of gas accretion into the Bondi radius then limits the accretion onto the planetary atmosphere, with a timescale of \citep{ADL2010}:\begin{align}
t_{\rm B} = \left(\mathcal{C}_{\rm B}\Omega\right)^{-1}\left(\frac{M_*}{a^2\Sigma}\right)\left(\frac{a}{H}\right)^{-7}\left(\frac{M_{\rm plnt}}{M_*}\right)^{-2},
\label{eq:gas03}
\end{align}
where $a$ is the orbital radius of the proto-planet, $H$ is the gas scale height, $\Sigma$ is the gas surface density, and $\mathcal{C}_{\rm B} \simeq 2.6$ is a constant meant to match this simple prescription to full 3D hydrodynamic simulations.

A second relevant length scale is the planetary Hill radius:\begin{align}
R_{\rm H} = a\left(\frac{M_{\rm plnt}}{3 M_*}\right)^{1/3},
\label{eq:gas04}
\end{align}
within which the planet's gravitational force exceeds the centripetal and gravitational force of the host star. Gas accreting into this radius is \citep{ADL2010}:\begin{align}
t_{\rm H} = 3\left(\mathcal{C}_{\rm H}\Omega\right)^{-1}\left(\frac{M_*}{a^2\Sigma}\right)\left(\frac{a}{H}\right)^{-1},
\label{eq:gas05}
\end{align}
where $\mathcal{C}_H \simeq 0.9$ to match the analytic presciption to numerical results. The accretion rate onto the planet will be described by the maximum of the three timescales:\begin{align}
t_{\rm acc} = {\rm max}\left(t_{\rm KH},t_{\rm B},t_{\rm H}\right),
\label{eq:gas06}
\end{align}
such that:\begin{align}
\frac{dM_{\rm plnt}}{dt} = \frac{M_{\rm plnt}}{t_{\rm acc}}.
\label{eq:gas07}
\end{align}

When the planet is embedded in the protoplanetary disk the region from which the planet will feed, is given by the smaller of the two length scales in equations \ref{eq:gas02} and \ref{eq:gas04}. At a certain point the planet's Hill radius exceeds the disk local scale height, and a gap in the disk is opened. Here the gas begins to primarily accrete from higher above the midplane as discussed earlier. We assume that the region of the disk from which the gas is accreted is between one and three scale heights.

The planet can open the gap in two ways: first, if the planet's Hill radius exceeds the local gas scale height then its gravitational influence will exceed the force from gas pressure. Second, the gap can open if the planet's gravitational force exceeds the viscous force of the surrounding gas disk along the midplane. Here we assume that once one of these criteria are met, then they are both met. The planetary mass repsonsible for either of these effects is \citep{MP03}:\begin{align}
M_{\rm gap} = M_* \min( 3h^3, \sqrt{40\alpha_{\rm turb} h^5} ),
\label{eq:gas08}
\end{align}
where $\alpha_{\rm turb}$ is the turbulent parameter from the standard $\alpha$-disk prescription \citep{SS73}, and $h = H/a$. For the purpose of the planet formation model we assume that $\alpha_{\rm turb} = 0.001$.

In principle, once the above requirement is met the gap is not completely devoid of gas \citep{Fung2014,Ginzburg2019,GyeolYun2019}, and often a cirumplanetary disk forms around the growing planet \citep[see for example,][]{Szul2014}. However for the purpose of further growth, the density inside the gap is not the relevant one to planet formation, since gas from the surrounding protoplanetary disk continues to flow into the gap \citep{Teague2019}. When the gap is opened we assume that the gas undergoes a meridionial flow, with a rate described by \cite{Morbidelli2014}. The rate of gas flux into the gap has the form of \citep{Cridland2018}:
\begin{align}
\mdot_{\rm gap} = 8\pi\nu(a/H)\Sigma_g,
\label{eq:gas09}
\end{align}
where $\Sigma_g$ is the unperturbed gas surface density. The gas viscosity has the form: $\nu = \alpha_{\rm turb}c_sH$ using the standard $\alpha$-prescription. The rate of gas accretion onto the planet is thought to be regulated by the dynamo-driven magnetic field generated by the growing planet \citep{Batygin2018}. The rate of accretion onto the planet given the impact of this large scale magnetic field was derived by \cite{Cridland2018}, which has the form:
\begin{align}
\frac{dM_{\rm plnt}}{dt} = \mdot_{\rm mag} = \frac{M_\oplus}{{\rm yr}}\frac{4}{3^{3/4}}\left(\frac{R_0}{R_H}\right)^2\left(\frac{M_{\rm plnt}}{M_\oplus}\right)^{-2/7}\left(\frac{\mdot_{\rm gap}}{M_\oplus/{\rm yr}}\right)^{3/7},
\end{align}
where $R_0 = \left(\pi/2\mu_0 \mathcal{M}^4/GM_\oplus^3/{\rm yr}\right)^{1/7}$\footnote{Note there was a typo in \cite{Cridland2018} that we have corrected here}, and $\mathcal{M} = BR_{\rm plnt}^3$ is the magnetic moment of the (assumed) magnetic dipole, with strength $B=500$ Gauss, and we assume that the size of the growing planet is a constant $R_{\rm plnt} = 2 R_J$. When the gap opens, the feeding zone shifts to between one and three scale heights \citep[as observed by][]{Teague2019}, and we assume that the rate of gas accretion is the smaller of $M_{\rm plnt}/t_{\rm KH}$ and $\mdot_{\rm mag}$.

\subsection{ Radial evolution through planetary migration }

As a planet grows it perturbs the surrounding gas, causing a transfer of angular momentum between the disk and planet. At low masses these perturbations come in the form of spiral waves emanating from Lindblad resonances, and gas on horseshoe orbits near the planet. Often the combined effect of these torques results in inward migration (known as Type-I, \citealt{GoldTrem79,W91,Paard14}). Recently however the details of Type-I migration have become more complicated, because disks may be less turbulent than previously thought \citep{BaiStone2016}, and the details of the gas temperature structure in the disk complicate the strength and net direction of angular momentum transport \citep{McNally2017,McNally2019,Crid19a,Guilera2019}.

Due to the current complexity of Type-I migration, we elect to ignore the impact of this low mass migration route. We keep the radii constant until the gap is open, at which point we begin to migrate the planet through Type-II migration \citep{LP86,LP93}. Given that our primary interest is studying what chemical impact the vertical accretion of gas will have, the particular migration prescription will not affect our experiment.

When the planet is sufficiently massive to open a gap, it suppresses the torques responsible for Type-I migration. At this point the planet acts as a mediator of angular momentum transport between the inner disk and the outer disk, moving angular momentum to the outer disk and migrating inward as a result. Type-II migration evolves on the viscous timescale: \begin{align}
t_{\nu} = \frac{a^2}{\nu}.
\label{eq:rad01}
\end{align}
The orbital radius of the planet evolves as: $da/dt = -a/t_\nu$.

We begin each synthetic planet at a different orbital radius $R_{\rm initial}$. The initial radii are distributed simply throughout the extend of the disk: R$_{\rm initial} \in [ 0.5, 1, 1.5]$, then every 1 AU from 2 to 20 AU, then every 2 AU up to 100 AU.

\subsection{ Chemical inheritance }

When it is embedded in the protoplanetary disk a growing planet has access to any gas that is under the influence of its gravity within the Hill radius, or the gas that is gravitationally bound within the Bondi radius. Equivalently, a planet's feeding zone is the smaller of its Hill and Bondi radii. Before the gap is opened, we assume that the number of carbon and oxygen atoms that are accreted into the planetary atmosphere is the average number of carbon and oxygen in the gas within the planet's feeding zone.

When the gap has opened (M$_{\rm plnt}$ $>$ M$_{\rm gap}$) we assume that the carbon and oxygen comes uniformly from between one and three scale heights. Even though the gas density drops with height, the gas velocity into the gap is higher at greater heights \citep{Teague2019} - hence we assume the mass flux of gas is constant with height above the midplane. Again we average the number of carbon and oxygen atoms between one and three scale heights to determine the incoming elemental abundances onto the planetary atmosphere. Furthermore we assume that no chemical change occurs during the vertical fall of the gas towards the planet, we justify this assumption in section \ref{sec:chemevo}.

\begin{figure}
\centering
\includegraphics[width=0.5\textwidth]{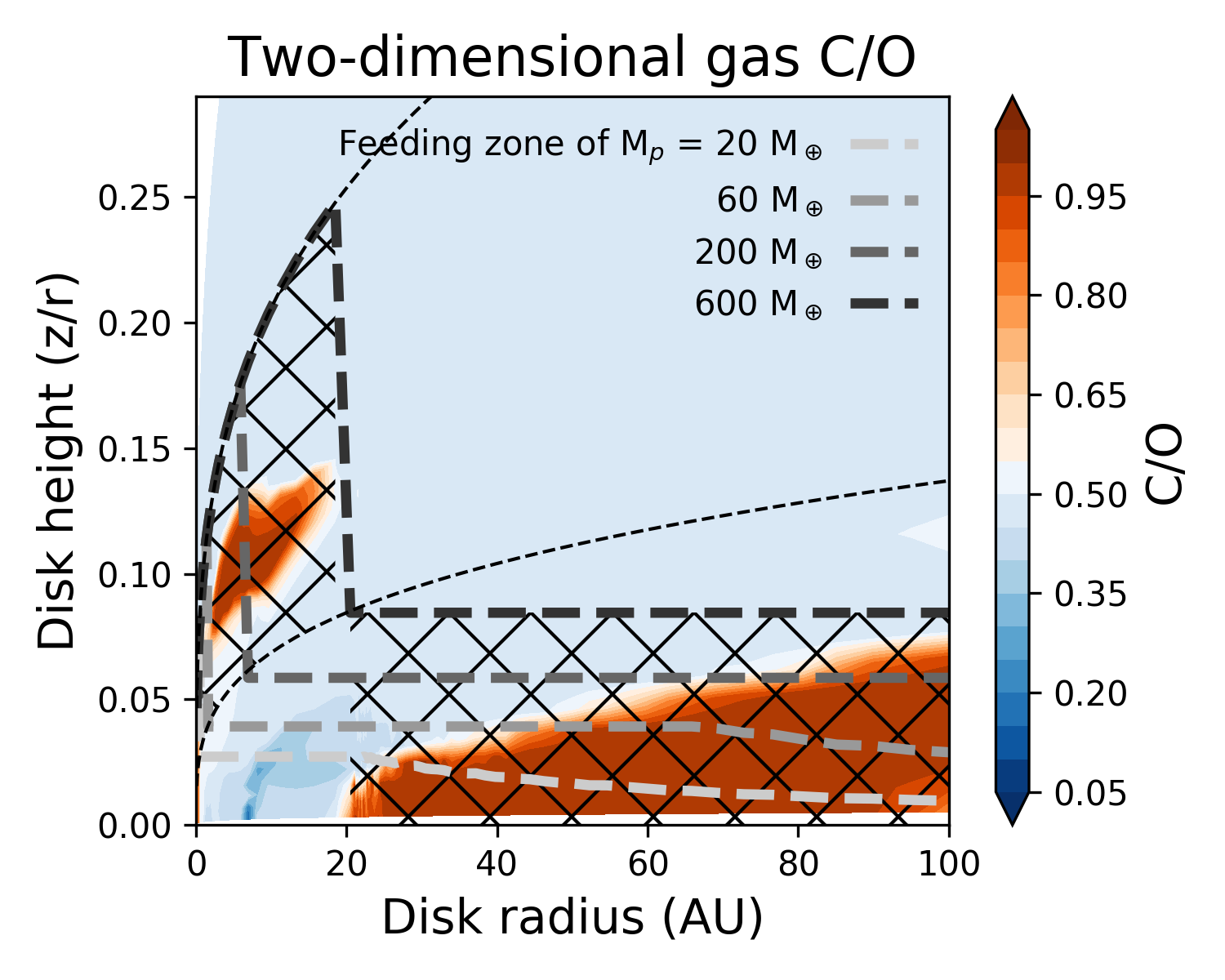}
\caption{The linear version of Figure \ref{fig:chemmodel02}, left panel, including the region of the disk within which the growing planet will accrete its gas - the `feeding zone' - for a given disk radius. The small dashed lines show the disk scale height and three times the disk scale hight. When the planet opens a gap (ie. the feeding zone exceeds the gas scale height), the planets begins feeding on gas accreted between the gas scale height and three times the gas scale height. We show the extent of the feeding zone for the 600 M$_\oplus$ planet with the black hatches. Clearly, once a planet opens a gap it changes the region of the disk from which the planet feeds.}
\label{fig:gasmodel01}
\end{figure}

In figure \ref{fig:gasmodel01} we show the C/O from figure \ref{fig:chemmodel02} with a linear x-axis along with the maximal height in the disk that contributes to the accretion of the planet, which is often known as the `feeding zone'. We show the feeding zone for a series of planets with different mass and at different orbital radii. In this way we see the difference that accretion from a few gas scale heights above the midplane could have on the resulting C/O of the growing planet. Above the midplane different chemical reactions dominate the partition of carbon and oxygen between gas and ice than along the midplane. Outward of 20 AU the shown planets have not opened a gap in the disk, however since the feeding zone grows past the extend of the CO ice surface for planet masses greater than 20 M$_\oplus$. Hence with this new prescription we will can still see some deviation in the 2D accretion model when compared to the 1D accretion model.

Once we know the average C/H and O/H that is accreted within a particular timestep $dt$ within the feeding zone of the planet, the flux of any element $X$ is:\begin{align}
\frac{dX}{dt} = \frac{1}{\mu m_H}\frac{dM_{\rm plnt}}{dt}\times (X/H)_{\rm gas},
\label{eq:gas09}
\end{align}
where $\mu m_H$ is the average weight of the gas particles. The micron sized dust grains are coupled to the gas, and hence as the gas is accreted, so too are the smallest grains. We again average the number of carbon and oxygen elements in the ice phase and scale the number by the fraction of small grains ($f_{\rm small}$) and the total number of hydrogen atoms in the gas. Hence including the micron grains gives:\begin{align}
\frac{dX}{dt} = \frac{1}{\mu m_H}\frac{dM_{\rm plnt}}{dt}\times \left[(X/H)_{\rm gas} + f_{\rm small}(X/H)_{\rm ice}\right].
\label{eq:gas10}
\end{align}
The micron grain fraction $f_{\rm small}$ is only a few percent along the midplane, but approaches unity rapidly near one scale height.

\begin{figure}
\centering
\includegraphics[width=0.5\textwidth]{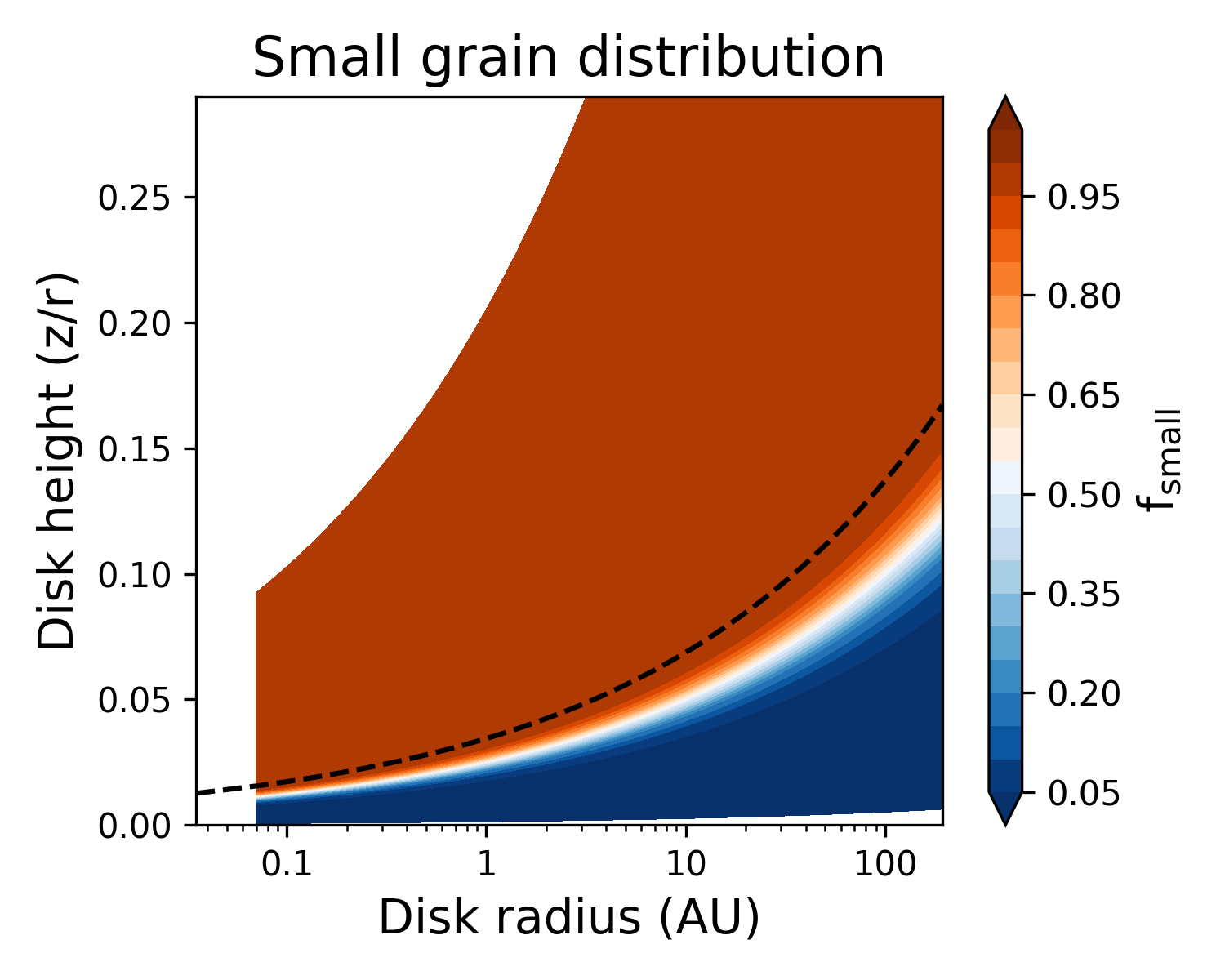}
\caption{Distribution of the small grain population. The small grains dominate the dust mass near the gas scale height (dashed line) and above.}
\label{fig:cheminh01}
\end{figure}

In Figure \ref{fig:cheminh01} we show the distribution of small grains ($f_{\rm small}$). Near the midplane the dust mass is dominated by the large grains which have efficiently settled, while nearer to the gas scale height (and above) the dust mass is dominated by the small grains. Since the small grains are coupled to the gas they can contribute (chemically) to the planetary atmosphere primarily in the two dimensional accretion case - after the planet has opened a gap. 

On the other hand, the large grains dominate the dust mass below the gas scale height, and act as the primary carrier of icy volatiles along the midplane. We neglect the accretion of large grains here since our embryos begin their growth near the pebble isolation mass ( with a nominal value of $\sim 20$ M$_\oplus$, \citealt{Bitsch2015,Bitsch2018}). Above this isolation mass, the planet has formed a positive pressure gradient in the disk gas that stops inflowing pebbles.

\begin{figure*}
\centering
\begin{overpic}[width=\textwidth]{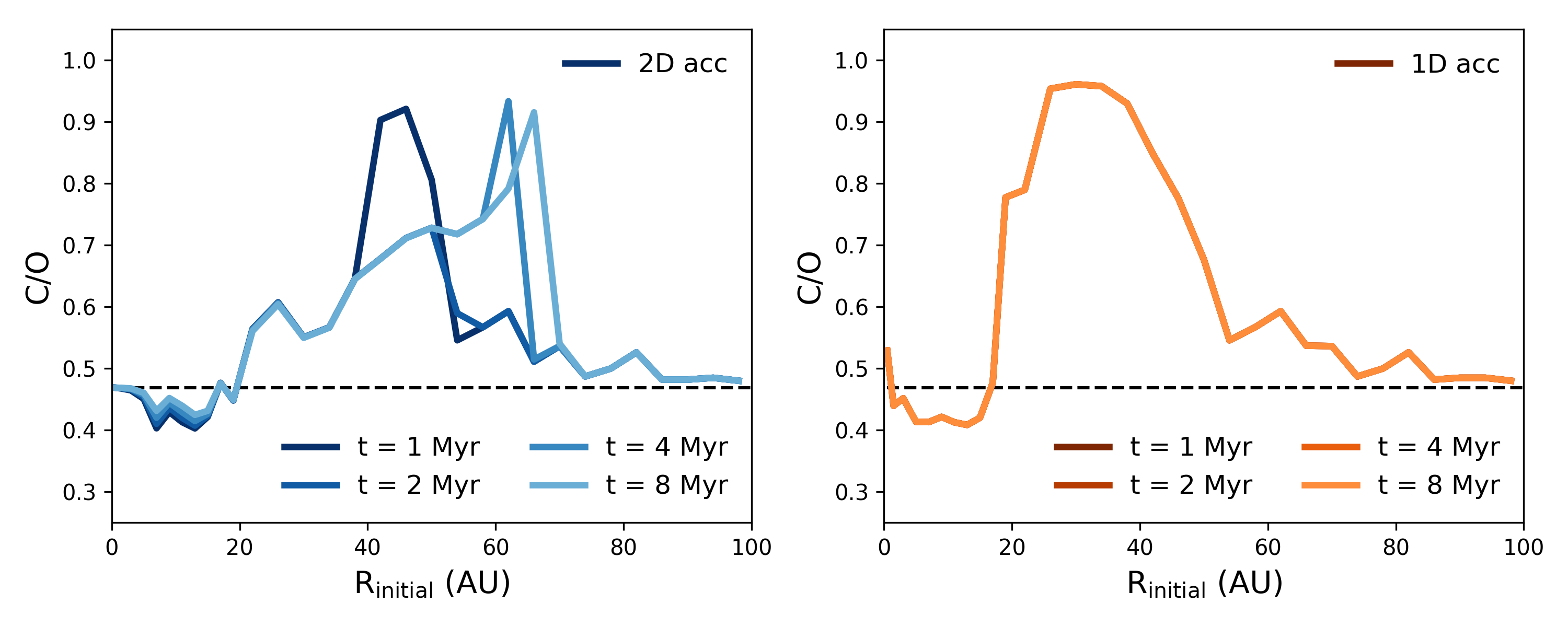}
\put(9,16){Disk volatile C/O}
\end{overpic}
\caption{The resulting atmospheric C/O for a range of planets with different starting radii. We neglect the orbital migration for these growing planets, hence they sample the gas at their position. The 2D accretion model is on the left, while the 1D accretion model is on the right - we show the evolution of atmospheric C/O using different shades of the colour. We note C/O of the disk volatiles before any chemistry partitions carbon and oxygen between gas and ice.}
\label{fig:results01}
\end{figure*}

\section{ Results: Differences in C/O }\label{sec:results}

\subsection{ Ignoring planetary migration }

Here we outline the results of changing the geometry of gas and small grain accretion into the atmosphere of growing planets. To simplify our analysis and interpretation, we first begin by simulating the growth of simulated atmospheres while ignoring the radial evolution of the proto-planet caused by migration.

In Figure \ref{fig:results01} we show the atmospheric C/O for a range of planets with different initial radii (R$_{\rm initial}$). In both the 1D and 2D accretion models we find that planets which start their evolution inward of 20 AU end with a smaller C/O than is initially in the gas (dashed line). This lower C/O is related to the chemical reactions discussed earlier which transforms gaseous CO to frozen CO$_2$.  Closer to the host star, near to the CO$_2$ ice line (at the midplane), the atmospheric C/O approaches the disk gas C/O. At larger radii than 20 AU the planets are growing outward of the CO ice line. As a result these planets initially accrete gas with low abundances of carbon and oxygen, and C/O close to unity. At very far radii (r $>$ 70 AU) the disk gas is again enhanced with volatiles by the photodesorption of the ice by the external UV field, which returns the atmospheric C/O to the solar value. 

Nearer to the edge of the CO ice surface the gaseous carbon content is marginally higher, and contributes more strongly to the atmospheric C/O. In the 1D accretion model planets between 20 and 40 AU quickly accrete an atmosphere with C/O near unity. While in the 2D accretion model, planets between 20 and 40 AU open a gap and accrete gas at heights with lower C/O which results in lower (but still super-stellar) C/O than in the 1D accretion model. 

\begin{figure}
\centering
\includegraphics[width=0.5\textwidth]{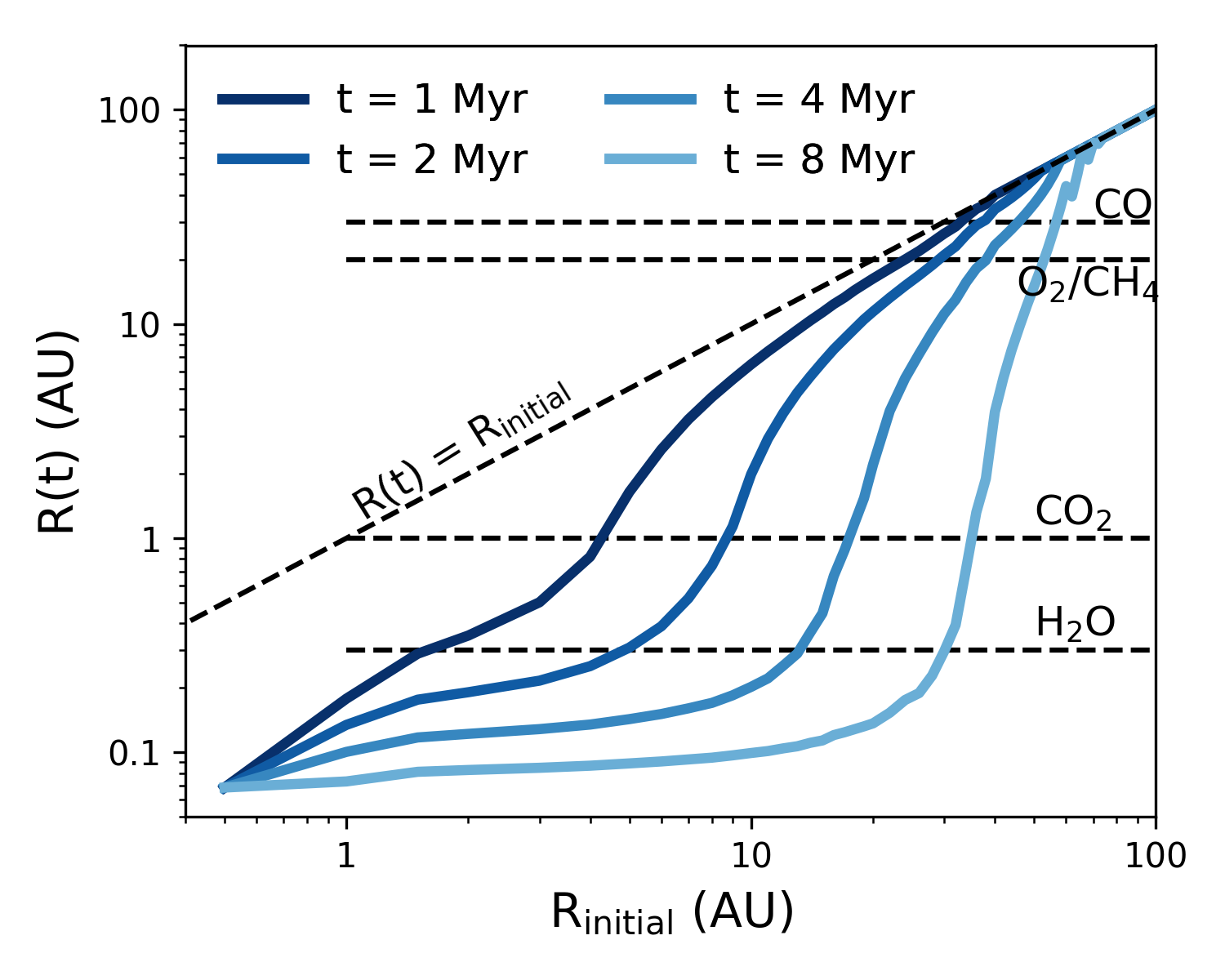}
\caption{Radius evolution of each synthetic planets (for both the 1D and 2D models). We note the locations of the midplane location of H$_2$O, CO$_2$, O$_2$, and CO ice lines as horizontal dashed lines for comparison. Note that the O$_2$ and CH$_4$ ice lines occur at similar positions, however in our chemical model O$_2$ is more abundant and hence its ice line is more relevant. Planets that begin within $\sim$15 AU migrate inward of the water ice line by 4 Myr and hence could accrete `pristine' gas. Inward of the water ice line, all volatiles are in the gas phase.}
\label{fig:results03}
\end{figure}

\begin{figure*}
\centering
\begin{overpic}[width=\textwidth]{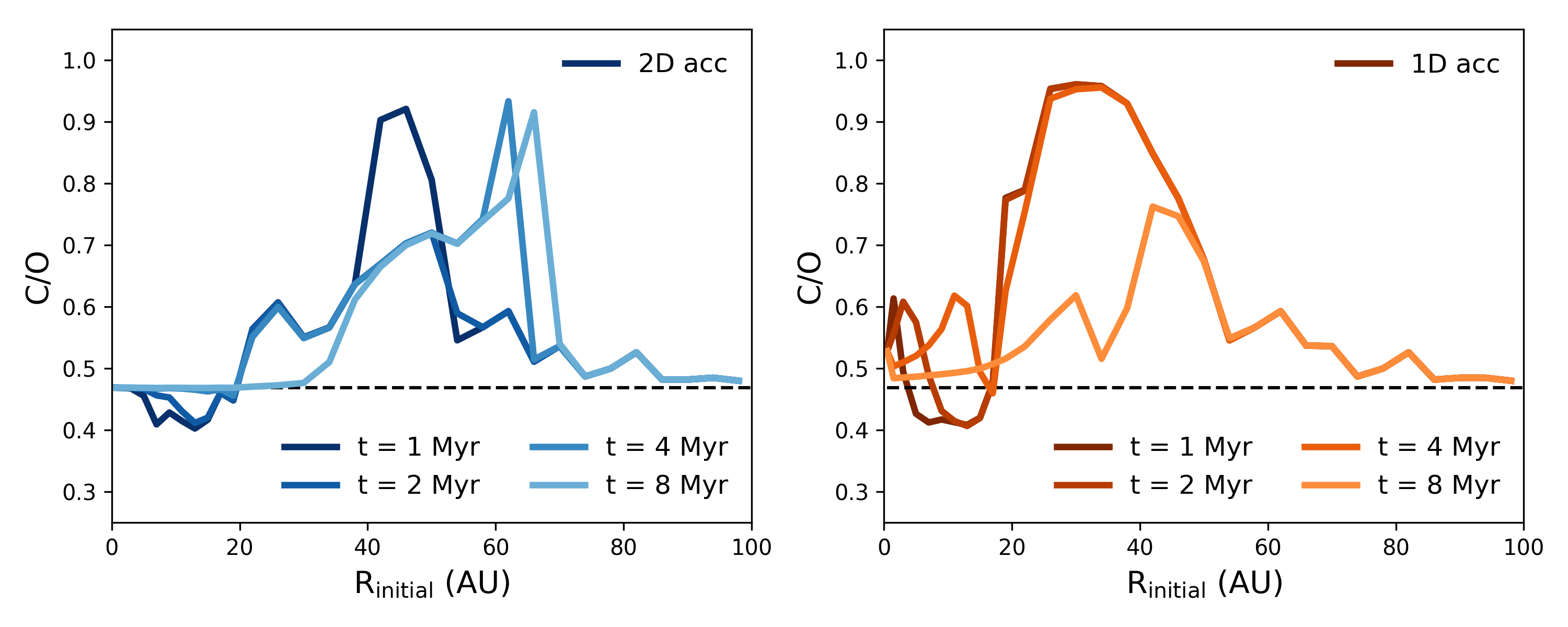}
\put(9,16){Disk volatile C/O}
\end{overpic}
\caption{The same as in Figure \ref{fig:results01}, but for the case where include orbital migration.}
\label{fig:results02}
\vspace{0.4cm}
\begin{overpic}[width=\textwidth]{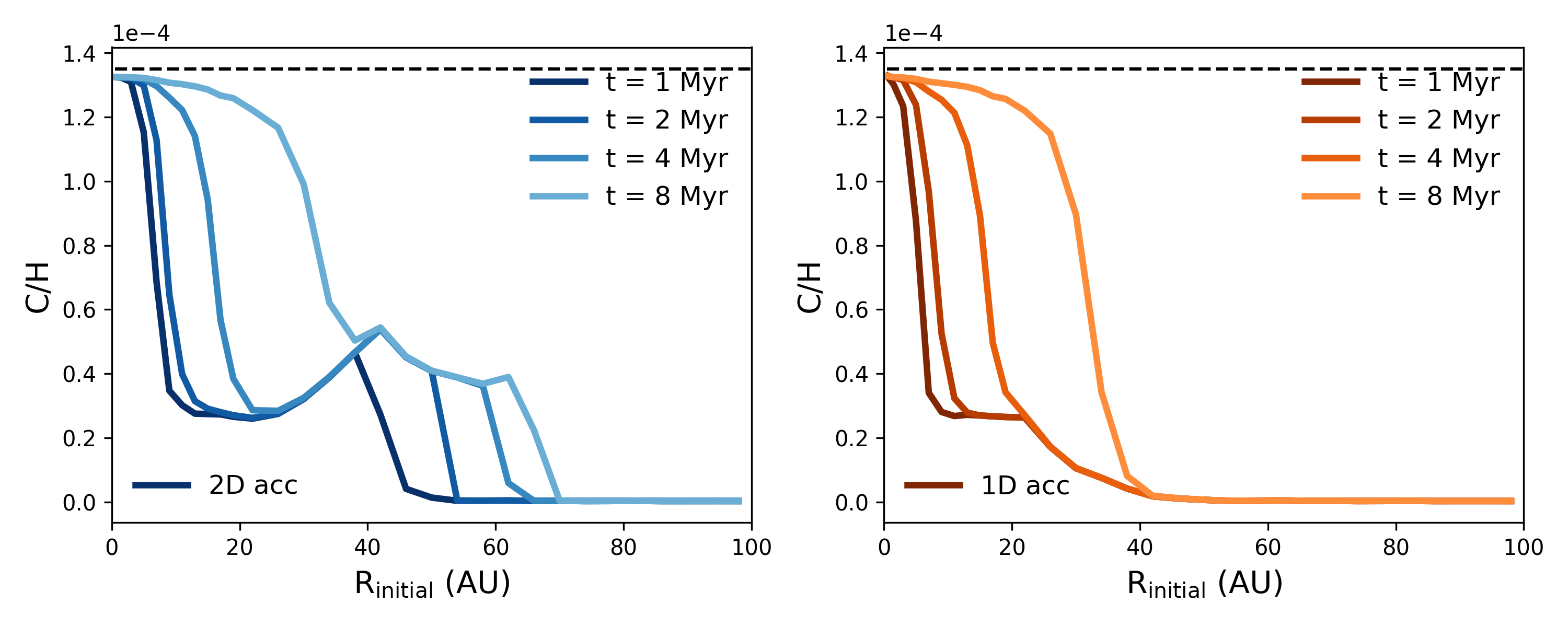}
\end{overpic}
\caption{The same as in Figure \ref{fig:results01}, but for the carbon-to-hydrogen ratio of each of the synthetic planets. The dashed line shows the disk's initial carbon to hydrogen ratio.}
\label{fig:results02xx}
\end{figure*}

The planets forming in the 1D accretion model see no evolution in their C/O since they can not sample the chemical properties of any other region of the disk. Likewise, planets in the 2D accretion model that do not open a gap in their disk ($R_{\rm initial}$ $>$ 60, see below) do not sample any material above the gas scale height, hence show minimal evolution in their C/O. An exception to this can be found for planets between 60 and 70 AU which show late stage (t $\ge$ 4 Myr) evolution in their C/O. While they do not open a gap in their disk, their feeding zone becomes sufficiently large that it exceeds the height of the CO snow surface, which allows for the accretion of CO rich gas. Since these planets originally accrete from carbon and oxygen poor gas (see Figure \ref{fig:chemmodel01}) this late stage accretion dominates the planet's C/O. For planets growing farther from their host star, the disk is more flared and the mass of the planet needed to feed from carbon and oxygen rich gas is larger. As a result this process occurs later in the disk lifetime, causing the peak in C/O to appear to move outward.

\subsection{ Including type-II migration }

We compute the orbital migration of each planet using Type-II migration discussed above. Migrating planets sample a wider range of the disk chemistry which could lead to further differences between the accretion models.

In Figure \ref{fig:results03} we show the evolution of the orbital migration of each of the synthetic planets related to the locations of abundant volatile midplane ice lines. All planets that started inward of $\sim$2 AU migrated inward of the water ice line very early ($t$ $\sim$1 Myr) in the disk life time. This gas is `pristine' because all of the volatiles are in the gas phase, and the grains are largely free of ice. 

By 4 Myr all planets which began within $\sim$15 AU have migrated inside the water ice line. By the same time, all planets beginning within $\sim$30 AU have migrated inside of the O$_2$ ice line. Migration is generally less efficient farther out in the disk because the density of gas is lower, weakening the torques on the planets.

In Figure \ref{fig:results02} we present the final C/O of the synthetic planets in the model which includes migration. We find that planets which began within $\sim$20 AU in the 2D accretion model show marginally higher C/O than is seen in the non-migrating models. This comes from the planets moving out of regions of low C/O between the CO$_2$ and O$_2$ ice surfaces to inward of the water ice line by 4 Myr (Figure \ref{fig:results03}). Hence these planets accrete gas that has the same C/O as was used as the initial condition for the chemical calculation.

Likewise, planets that begin within $\sim$40 AU move across the O$_2$ ice line into a region of low gaseous C/O ratio, reducing their final atmospheric C/O by 8 Myr. Outward of 40 AU the planets have not migrated enough to move into a region of the disk with a different chemical structure from where it began, and has little impact on the final chemical properties of the planetary atmospheres. This implies that the composition of directly imaged planets may be a good measure of the local chemical properties of their natal gas disks. 

In Figure \ref{fig:results02xx} we present the carbon-to-hydrogen ratio (C/H) of the synthetic planets shown in Figure \ref{fig:results02}. We note that planets which formed inward of 20 AU show nearly equal atmospheric C/H to that of the disk, which is caused by early gap opening and the accretion of gas and icy dust grains that are not depleted in carbon. 

By contrast, planets outward of 20 AU take longer to open their gap, and spend a significant fraction of their lifetime accreting gas and small grains that are depleted in volatile carbon. In those regions of the disk the volatile carbon is depleted on the large dust grains, whose accretion we ignore in this work. We note that while our synthetic planets are all sub-solar in their C/H, our own Jupiter has super-solar abundances of C/H \citep{Atreya2016}. This discrepancy suggests that the accretion of planetesimals is important to the final carbon abundance of planetary atmospheres since the accretion of pebbles (the large grain population in our work) is suppressed by gap opening. We discuss the implication of the difference between Jupiter and our synthetic planets below.

\begin{figure*}
\centering
\subfigure[ Results for the 2D accretion model  ]{
\includegraphics[width=0.5\textwidth]{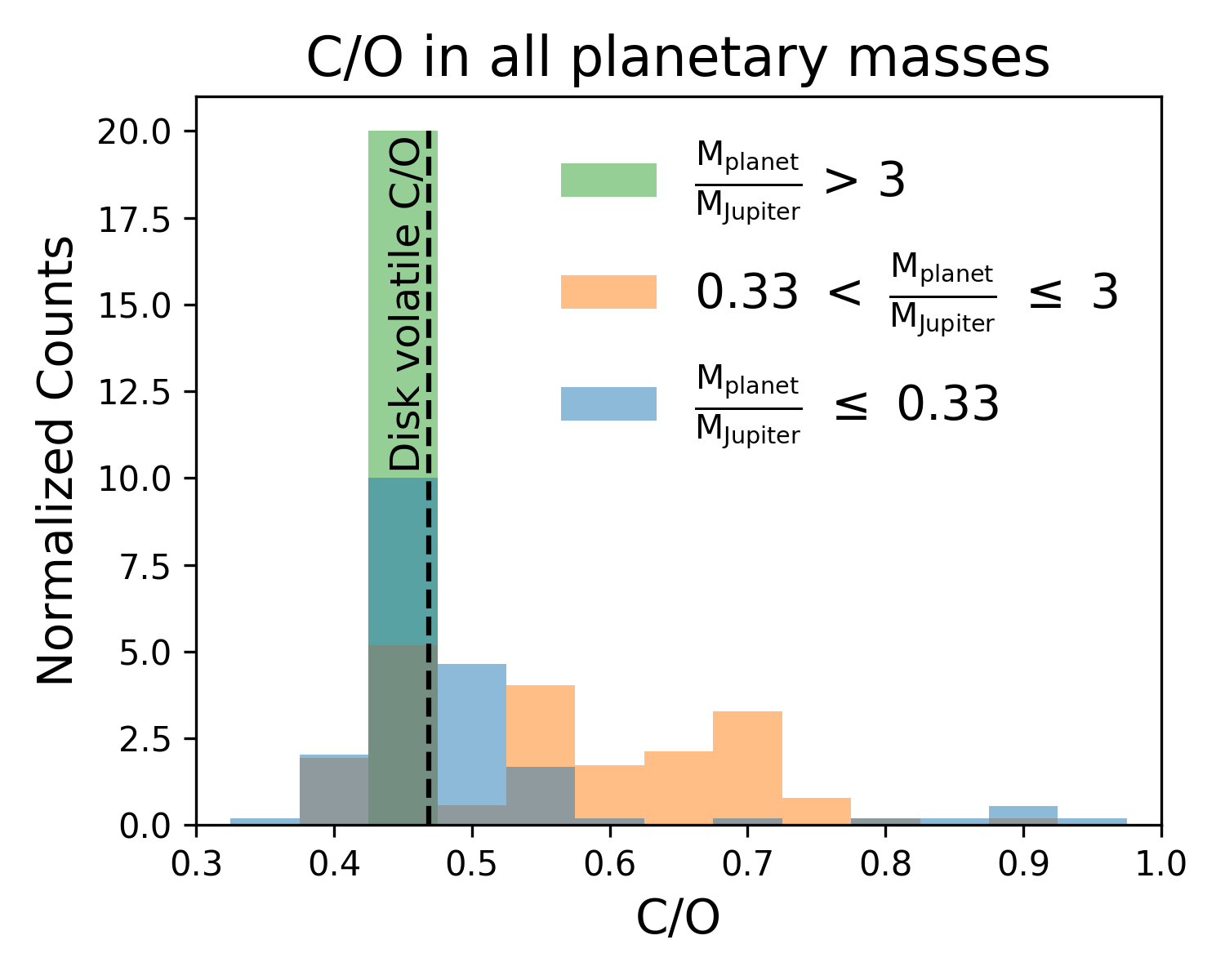}
\label{fig:results04}
}
\subfigure[ Results for the 1D accretion model ]{
\includegraphics[width=0.5\textwidth]{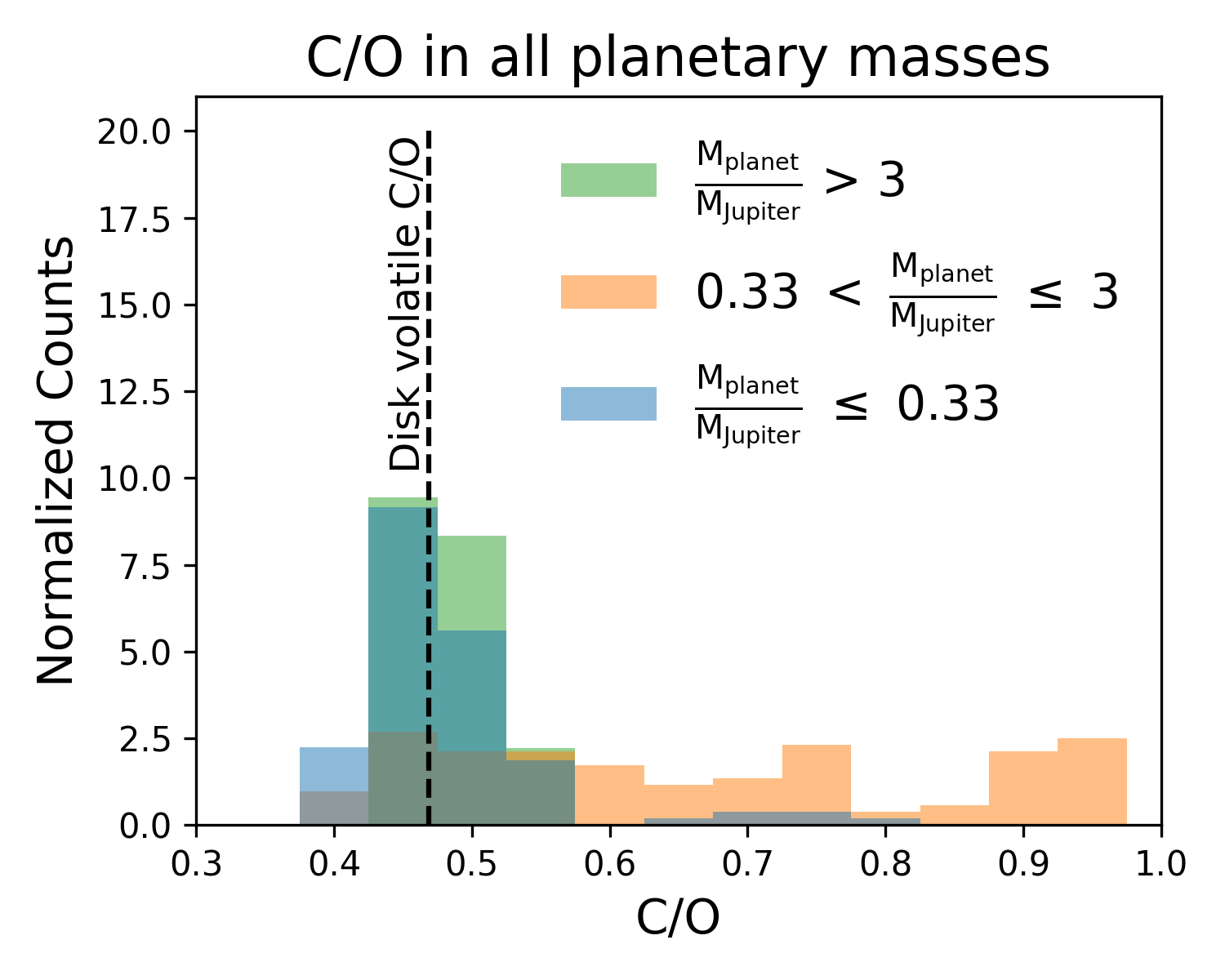}
\label{fig:results04b}
}
\caption{A histogram of C/O for different planetary masses throughout the history of planet formation for each synthetic planet. We normalize each histogram such that the integral is equal to unity. We note C/O for the all volatiles (gas and ice) in the disk for comparison with a dashed line. Note that for the high mass planets in the bin which includes the disk volatile C/O (dashed line), the majority of planets have exactly the same C/O as the disk (recall the inner 30 AU at 8 Myr of Figure \ref{fig:results02}). }
\label{fig:results04f}
\end{figure*}

We noted earlier that our disk model was meant to model a presolar system, which describes the evolution of pre-main sequence, low-mass stars. Higher mass stars (with masses of a few M$_\odot$) have been found to host the majority of directly imaged exoplanets. The pre-main sequence phase of these higher mass stars are known as Herbig Ae systems, and are typically hotter, more irradiated systems. The hotter gas would predominately push the volatile ice lines farther away from the host star. Since the relation of C/O to R$_{\rm initial}$ depends on the location of the ice lines, then in a Herbig Ae disk the curves in figures \ref{fig:results02} and \ref{fig:results02xx} would shift to the right.

\section{ Discussion: Implications of vertical accretion }\label{sec:disc01}

Given that planet migration appears to be a natural consequence of a planet interacting with its protoplanetary disk in numerical simulations, we view this model to be more physically motivated than the non-migrating model. In what follows we will discuss only the results connected to the migrating formation model.

\subsection{ Mass dependence of C/O }

Here we have studied the chemical impact of different feeding zones on the atmospheres of exoplanets. We elect to initiate our formation model from 10 M$_\oplus$ which ignores the initial build up of the core, and any chemical impact that the core might have on the atmosphere. For Jupiter masses (and above) this assumption ignores core erosion and subsequent mixing of core material. This process is complex and depends strongly on the equation of state of the gas at high temperatures and pressures - which is not well constrained. At low masses this could become important since the mixing length between the core and atmosphere is smaller.

In Figure \ref{fig:results04} we show a normalized histogram of C/O for three different planetary mass ranges. These mass bins are: 1) sub-Saturns, 2) Saturn and Jupiter-like planets, and 3) super-Jupiters. We draw planet masses and their C/O at 1, 2, 4, and 8 Myr throughout the growth of the synthetic planets and combine them into a single histogram.

While a few sub-Saturn planets start outward of the CO ice line and hence do show high C/O nearing unity, generally we find that sub-Saturns are most similar with the disk gas C/O. As time passes some of these sub-Saturns grow in mass and become Saturn or Jupiter-like planets. This mass range has a much broader distribution, since some planets accrete more carbon-rich gas at large radii while some accrete oxygen-rich gas at smaller radii - this is further illustrated below.

The super-Jupiter planets have all evolved to have the same C/O as the initial disk gas. This is due to the fact that these planet originate within 20 AU (see below), and hence accrete the oxygen-rich gas and ice that is interior to the CO ice line. Similarly, these planets are massive enough to open a gap and vertically accrete, bringing in carbon and oxygen that is contained in the icy, micron sized grains.

In Figure \ref{fig:results04b} we see a similar distribution in the sub-Saturn population of C/O in the 1D accretion model. A primary difference for the Saturn and Jupiter-like planets is that their distribution in C/O is wider than in the 2D accretion model, allowing for generally higher C/O in the 1D accretion model. Similarly we find that the super-Jupiters have a slightly wider distribution in C/O. This comes from the fact that in the 1D accretion model, the planets have less access to the icy micron sized dust grains that populate near one disk scale height.

\subsection{ Observability of vertical accretion }

Currently our knowledge of the chemical composition of the planet forming region is limited since our primary gas tracers (CO and its isotopologues) can not trace down to the midplane of the disk. Instead these tracers are best used to probe above the gas scale height, in a region that is often called the `warm molecular layer'. This layer is warm enough and has high enough UV fluxes that gas chemistry can proceed more efficiently, and molecules stay in the gas phase more readily than along the midplane. Very rare (and optically thin) isotopologues like $^{13}$C$^{17}$O could be used to pierce the veil of the warm molecular layer (see for example \citealt{BoothA2019}) to probe closer to the midplane, however its rarity makes this difficult.

If it is true that gas falls vertically into the gap towards the planet then chemical characterization of the planet forming material becomes much easier. In this work we posit, based on the recent ALMA observations \citep{Teague2019}, that when the gap is opened the gas accretes primarily from between one and three scale heights. These regions will be available to probe with the \textit{James Webb Space Telescope} (\textit{JWST}).

\begin{figure}
\centering
\includegraphics[width=0.5\textwidth]{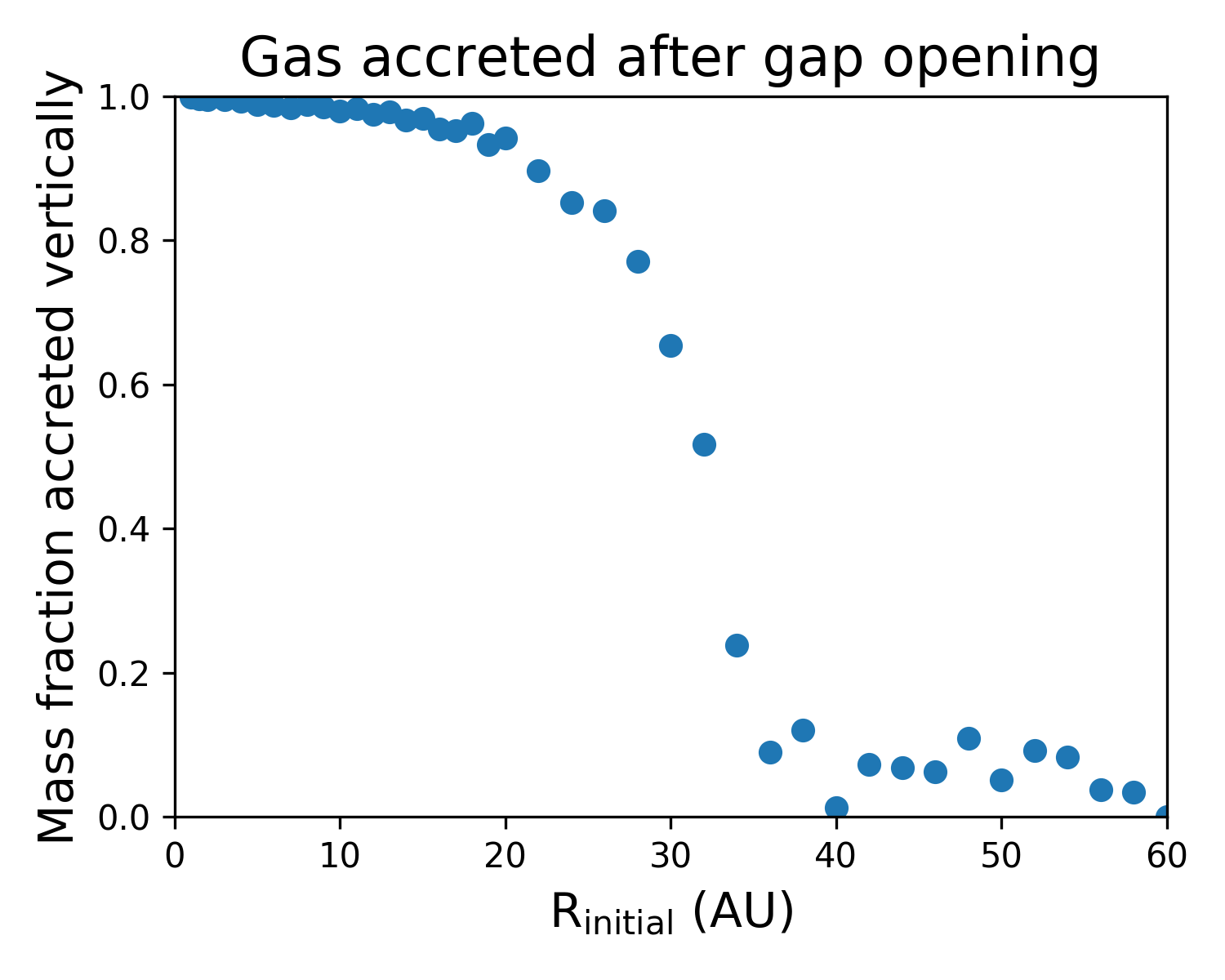}
\caption{The fraction of gas accreted after the planet gas opened a gap. Planets starting outward of 60 AU never open a gap and are hence left off the figure.}
\label{fig:results07}
\end{figure}

In Figure \ref{fig:results07} we show the fraction of gas that is accreted onto the planet after the gap has been opened in the disk. For planets starting inward of 20 AU the majority (90\%) of the gas mass is accreted after the gap has opened, when the vertical accretion of disk gas into the gap dominates. Inward of 30 AU, roughly half of the mass is accreted after the gap has been opened.

\ignore{
\begin{figure}
\centering
\includegraphics[width=0.5\textwidth]{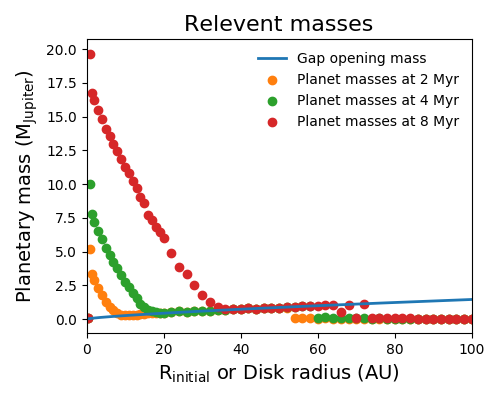}
\caption{ Evolution of the synthetic planet masses compared to the gap opening mass in the disk model. }
\label{fig:results09}
\end{figure}
}

By contrast, outside of 40 AU, less than 20\% of their mass is accreted after they open a gap. Not shown in the figure, outside of 60 AU, the gap is never opened within the 8 Myr lifetime of the disk. The dust gaps recently interpreted as embedded planets by \cite{Zhang2018} (ie. DSHARP VII) rely on planet masses of at least a Neptune mass at radii greater than 10 AU. More than half of their dust gaps are at disk radii of at least 50 AU, which is a region of the disk where our synthetic planets tend to not open gaps in the gas disk. While this difference seems difficult to reconcile, we note that the gas gap opening mass is generally larger than the mass needed to open a gap in the dust.

\begin{figure}
\centering
\includegraphics[width=0.5\textwidth]{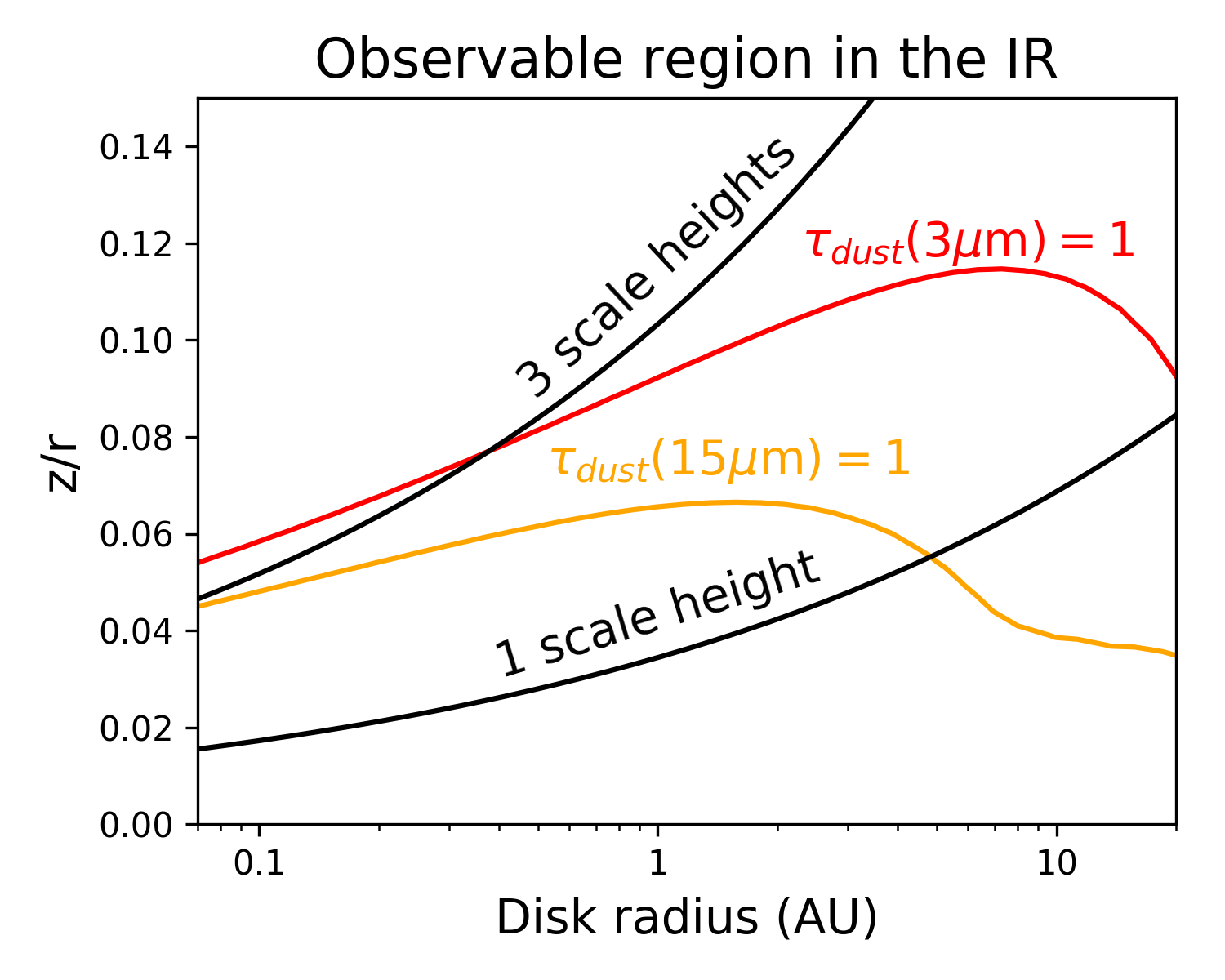}
\caption{ Contour of optical depth of unity for dust continuum at two relevant infrared (IR) wavelengths. At higher optical depths the dust will shroud the emission of warm molecular species. Near 15 $\mu$m molecule like CO$_2$, C$_2$H$_2$, and HCN have strong molecular emission features, while near 3 $\mu$m H$_2$O and CO dominate the emission spectra of the gas disk. }
\label{fig:results09x}
\end{figure}

In Figure \ref{fig:results09x} we show a schematic of the observability of gas molecular emission between one and three scale heights at infrared (IR) wavelengths. Only when the dust optical depth is lower than unity can the molecular emission of gas species can be seen. Near 15 $\mu$m are strong molecular features of H$_2$O,CO$_2$, C$_2$H$_2$, and HCN, while near 3 $\mu$m the more relevant molecules are H$_2$O and CO. All of these molecular species are relevant for estimating the C/O of the disk gas which will help to understand the chemical properties of the gas that accretes onto forming planets. The gas-to-dust ratio in the surface layers of our disk is 10000 and hence the contours in the figure represent the maximum depth that would be available to observations. For lower gas-to-dust ratios the dust opacity will be higher and the depth probed by IR will be shallower.

Assuming that the flux of gas entering into the gap from between one and three scale heights (as we assume in this work) then studying the IR emission of protoplanetary disks will probe the material that builds planets down to $\sim1$ AU. On the other hand, if the flux of gas is higher closer to one scale height rather than three then gas emission from between a few AU and 20 AU will be more reliably probed by IR observations of the inner disk.

From this we can say that studying the inner 10 - 20 AU of the disk at IR wavelengths can potentially teach us about the chemical properties of the gas that will build the Jupiter-size planets which migrate to near 1 AU. We thus require the sensitivity of \textit{JWST} to study the warm gas in the molecular layer at radii within 20 AU to understand the chemical properties of giant planets in exoplanet systems which harbour giant planets inward of 10 AU. At larger separation, the high angular resolution of the Atacama Large (sub)-Millimeter Array (ALMA) will continue to be important to learn about the chemistry feeding of planet formation that occurs far from the host star.

\subsection{ Chemical evolution during meridional circulation }\label{sec:chemevo}

A secondary effect of the described circulation is the possibility that the gas chemically evolves with time as it circulates. There are four relevant timescales to consider: (1) the dynamical timescale which describes the rate at which gas falls to the midplane after it crosses into the gap. (2) the timescale associated with the viscous evolution of the gas as it circulates back into the surrounding disk. (3) the diffusion time that is required for the proto-planetary disk to return to a hydrostatic equilibrium once gas is cycled from the circumplanetary disk to the proto-planetary disk and (4) the chemical timescale. 

The fourth timescale is related to the slowest chemical reaction in the system. The chemical timescale depends on the flux of UV radiation that is responsible for the photodissociation of molecular species. Roughly speaking this timescale is $1/t_{\rm chem} = G_0\times 10^{-10} s^{-1}$ \citep{Heays2017}, where $G_0$ parametrizes the strength of the radiation field relative to the interstellar radiation field ($G_0\sim 1$). When the chemical time is shorter than the physical timescale, we can say that chemical evolution is important to the chemical properties of the gas.

\begin{figure}
\centering
\includegraphics[width=0.5\textwidth]{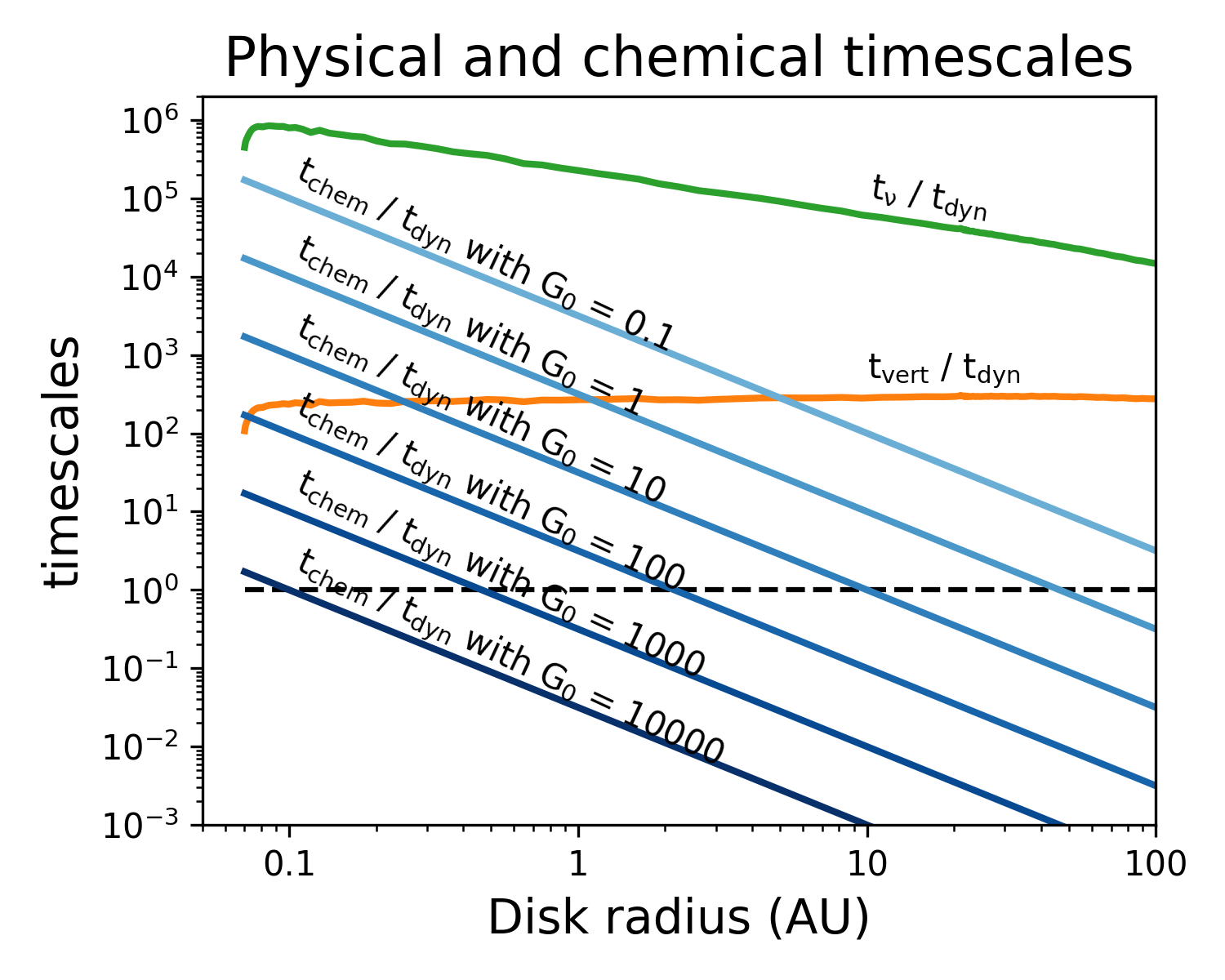}
\caption{Relevant physical and chemical timescales relative to the local dynamical timescale. We note that the chemical timescale scales with the local UV flux (parameterized by G$_0$) - higher UV flux leads to faster chemical timescales. For the standard interstellar UV flux (G$_0$ = 1), the chemical timescale is smaller than the vertical mixing timescale only outside 30 AU.  }
\label{fig:results08}
\end{figure}

In Figure \ref{fig:results08} we show the relevant timescales related to the circulation relative to the dynamical timescale. The diffusion timescale scales with the gas scale height and the gas diffusion: t$_{\rm vert}\sim$ $H^2/$D, where D $=\nu/Sc$ and $Sc\sim 1$, while t$_{\rm dym} = 2\pi/\Omega$ is the dynamical timescale that describes the  vertical fall from three scale heights. This timescale is the fastest for all radii, for G$_0 < 10000$ and hence all other timescales are plotted relative to the dynamical timescale. In addition, we can assume that the gas does not have time to chemically evolve as it falls from the disk towards the planet. We find that for the standard UV dissociation rate (G$_0 = 1$), the vertical mixing is faster than the chemical timescale only inside 1 AU. However, between one and three scale heights the gas in our disk model is irradiated with a UV flux consistent with G$_0 = 1000$. At this rate, we see that the vertical mixing is slower than the local chemical evolution, and hence we would expect the gas to be chemically changed as it diffuses back up from the midplane. 

Given that the highest UV flux occurs higher up in the disk atmospheres, we would expect that the circulating gas will chemically evolve towards the UV irradiated gas near three scale heights. To test whether this tendency would have any effect on our results we modified the feeding zone of the planets from between one and three scale heights to between 2.9 and 3 scale heights. This test showed only minor changes to the resulting atmospheric C/O, and hence we ignore the effect of chemical processing during meridonial circulation for the remainder of this work.

\subsection{ Carbon abundance of Jupiter-like planets }

As seen in Figure \ref{fig:results02xx}, all of our synthetic planets have lower carbon abundance (relative to hydrogen) than is initially in the gas disk. Our own Jupiter has super-solar C/H which is generally attributed to the accretion of planetesimals after its atmosphere had been established \citep{Atreya2016}.

\begin{figure}
\centering
\includegraphics[width=0.5\textwidth]{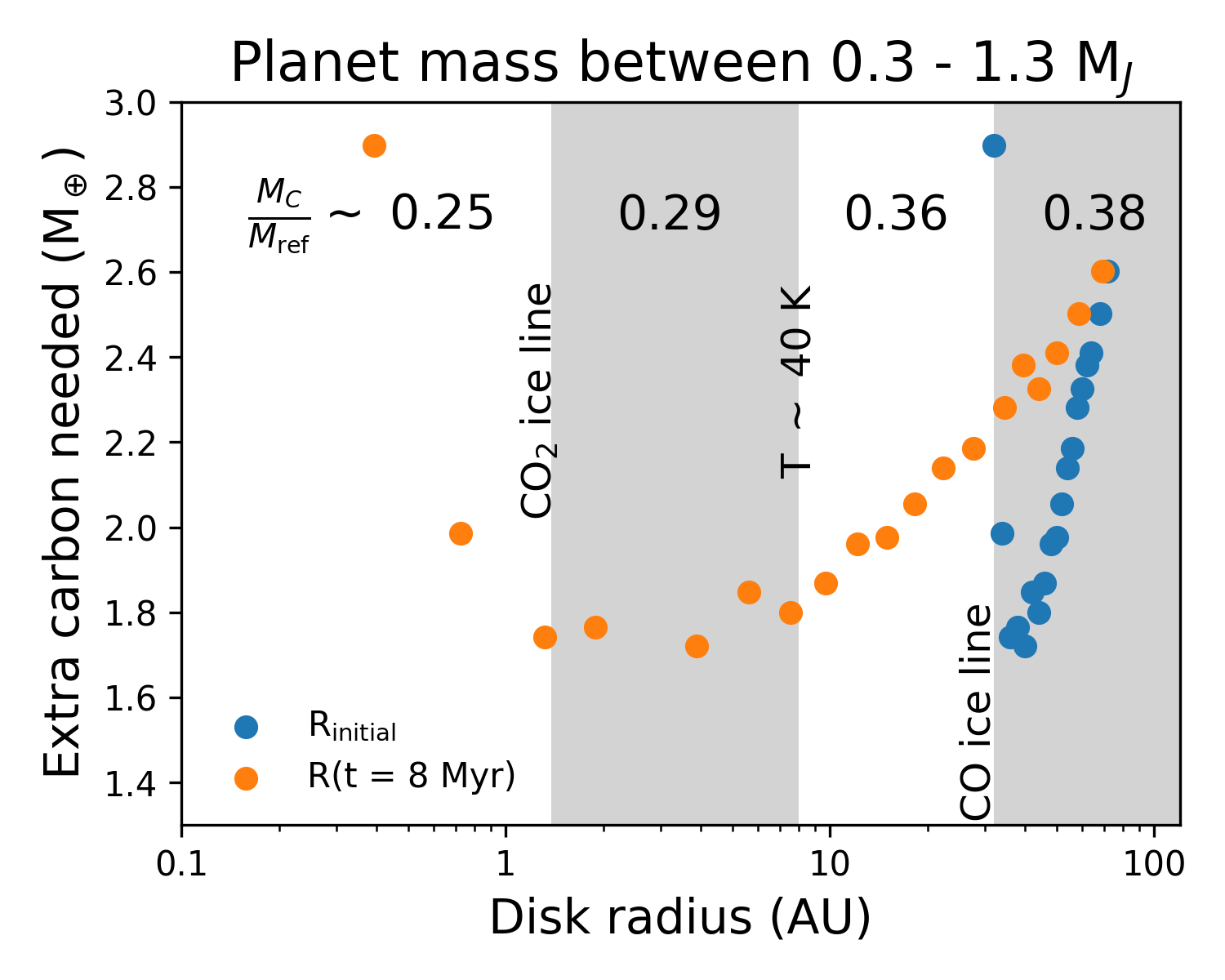}
\caption{ The mass of solid carbon needed to accrete into the atmosphere to bring C/H up to $\sim 4\times$C/H$_{\rm solar}$. In addition we show the mass fraction of carbon relative to the refractories (M$_C$/M$_{\rm ref}$) for the solids within the CO$_2$ ice line (r $<$ 1.4 AU), between the CO$_2$ ice line and the point where CO is converted to CO$_2$ ice (at T$_{\rm gas}$ $\lesssim$ 40 K, 1.4 AU $<$ r $<$ 8 AU), from there to the CO ice line (8 AU $<$ r $<$ 32 AU), and outward of the CO ice line (r $>$ 32 AU). }
\label{fig:results09}
\end{figure}

In Figure \ref{fig:results09} we compute the quantity of solid (ice and refractory) carbon needed to accrete in order to match the super-solar C/H seen in Jupiter ($\sim 4\times$ C/H$_{\rm solar}$) for planets between a Saturn and a super-Jupiter mass. We note first that all of these planets come from outward of about 30 AU and migrate inward to orbital radii akin to our solar system. We predict the solid carbon abundance in the planetesimals by assuming that they have the same mass ratio as in \cite{Mordasini16} - 2:4:3 by mass of carbon, silicates, and iron respectively. We then add any frozen species, which is dominated by CO$_2$ and CO - hence the solid carbon mass changes at the CO$_2$ ice line, where the dust temperature is below which CO gas is efficiently converted to CO$_2$ ice (T$\sim$40 K), and the CO ice line. We note the radial extent of these regions with the white and grey patches.

To create a Jupiter analogue (Jupiter mass at orbital radii of a few AU), roughly 1.75 M$_\oplus$ of solid carbon must be accreted. Assuming that this solid carbon is accreted only from planetesimals near its orbital radii then the solid mass that is required to be accreted is: $1.75$ M$_\oplus$ / 0.29 $\sim$ 6 M$_\oplus$. This mass is on the low end of what is required to explain the observed dust population in debris disks \citep{Krivov2018}, and more than likely our young solar system had more mass in planetesimals soon after the gas in the disk evaporated.

We note in Figure \ref{fig:results07} that planets which begin their evolution outward of 30 AU only accrete between 10-20 \% of their mass after the gap is opened. Hence they may similarly stay below their pebble isolation mass for longer than planets growing closer to the host star. In this way they may increase the metallicity of their atmosphere by accreting pebbles directly from the disk rather than through planetesimal accretion. Given the efficiency of pebble accretion, and the fact that there are plenty of pebbles leftover in the Class-II phase of protoplanetary disks \citep{Ansdell2016,Ansdell2017}, we could expect that the increase of C/H is feasible with pebble accretion.

Giant planets inward of our Jupiter analogue will require more solid mass accretion to match Jupiter's C/H, while planets ending outward of our Jupiter analogues would require less solid accretion. Along with the accretion of carbon, solid accretion will bring oxygen into the atmosphere of giant planets. In the 4 regions outlined in Figure \ref{fig:results09} the oxygen mass fraction (relative to total refractory mass) is 0.25, 0.32, 0.58, and 0.62 inward of the CO$_2$ ice line, inward of 8 AU, inward of the CO ice line, and outward of the CO ice line respectively. With that, the extra mass of oxygen accreted by our Jupiter analogue is 6 M$_\oplus$ $\times$ 0.32 = 1.92 M$_\oplus$. We would predict that our own Jupiter has twice the solar O/H, and thus a super-solar C/O due to its C/H enhancement by a factor of four.

These estimations rely on the accuracy of our chemical model which computes the settling of dust grains prior to computing the abundance of ice on the grains. In reality the gas chemistry and  dynamics occur concurrently and hence the ice abundance (relative to silicon) along the midplane could be enhanced by the growth and settling of ice-rich dust grains. \cite{Krijt2016} show that through dynamical evolution alone, the ice abundance would vertically homogenize. This mixing would in most cases enhance the ice abundance on the midplane, offering more volatile carbon and oxygen in the early stages of atmosphere accretion. Hence our estimated mass fractions represent the minimum carbon mass (relative to refractory mass) locked up in the solids.

\section{ Conclusions: Does vertical gas accretion impact the atmospheric C/O? }\label{sec:conclusion}

Our primary question centred on whether including vertical gas accretion would have any impact on the chemical properties of planetary atmospheres. Throughout this work we have compared the resulting atmospheric C/O between synthetic planets accreting gas from the midplane (1D accretion model) and from between one and three scale heights after the gap has opened (2D accretion).

We indeed find that including vertical accretion from higher in the disk typically results in planets with at least the disk gas C/O, and generally reduces the number of planets with C/O near unity when compare to the 1D accretion model. This reduction comes from the availability of oxygen rich icy grains that exist at heights above one scale height. Some of these differences are suppressed because enough time has elapsed and the planets migrate via Type-II migration through a wide range of different regions of the disk. Given that migration is a natural consequence of embedding a planet within a gaseous disk, we expect that long-lived ($\sim 8$ Myr) disks would tend not to produce planets that show the chemical alteration through 2D accretion - these long lived disks however are rare \citep{HLL01,Her07}.

For more moderate disk ages ($\lesssim$ 4 Myr) the chemical differences of 2D accretion is maintained for planets which begin forming between 20 and 40 AU - this includes planets that turn out to be Jupiter-analogues. Planets forming inward of 20 AU show smaller C/O relative to the disk in the 2D accretion model, while in the 1D accretion model planets tended to have larger C/O. This difference is small, with a shift in C/O of a maximum of 0.1-0.2, and it remains to be seen whether future medium and high-resolution spectrographs can accurately predict atmospheric C/O down to this accuracy.

We note that there is enough time during one part of meridonial circulation to chemically alter the gas. Hence we expect that the chemical structure of the gas will equilibrate to the chemical properties at the top of the circulation ($\sim 3$ H). This would require chemical modelling to confirm our hypothesis, however this is beyond the scope of this work.

While the changes may be small between 1D and 2D accretion, an important implication of this work is that we expect gas accretion to be dominated by vertical accretion for planets inward of 20 AU. Hence we can characterize planet forming material observationally more simply than if planets grow purely from gas on the midplane. We reiterate that the warm gas that is between one and three scale heights in the inner (r $<$ 20 AU) disk can be probed and chemically characterized with \textit{JWST}.

Given our knowledge of the amount of carbon accreted by our Jupiter-analogues we compute the required pebble or planetesimal accretion needed to enhance their atmospheres to the super-solar abundance of C/H. We find that we need at least 6 M$_\oplus$ of solid accretion to enhance our Jupiter-analogues. This mass is on the low end of the inferred solid mass of debris disks, and predicts a marginally (factor of 2) super-solar abundance of O/H in Jupiter.

\section*{ Acknowledgements }

We greatly thank Richard Teague for his lengthy discussions with both A.C. and A.B. preceding this work. We thank the anonymous referee for their helpful comments that improved the clarity of the text. Astrochemistry in Leiden is supported by the European Union A-ERC grant 291141 CHEMPLAN, by the Netherlands Research School for Astronomy (NOVA), and by a Royal Netherlands Academy of Arts and Sciences (KNAW) professor prize.

\bibliographystyle{aa} 
\bibliography{mybib.bib} 

\appendix

\section{ Two dimensional distribution of a few molecular species }\label{sec:app01}

\begin{figure*}
\centering
\includegraphics[width=\textwidth]{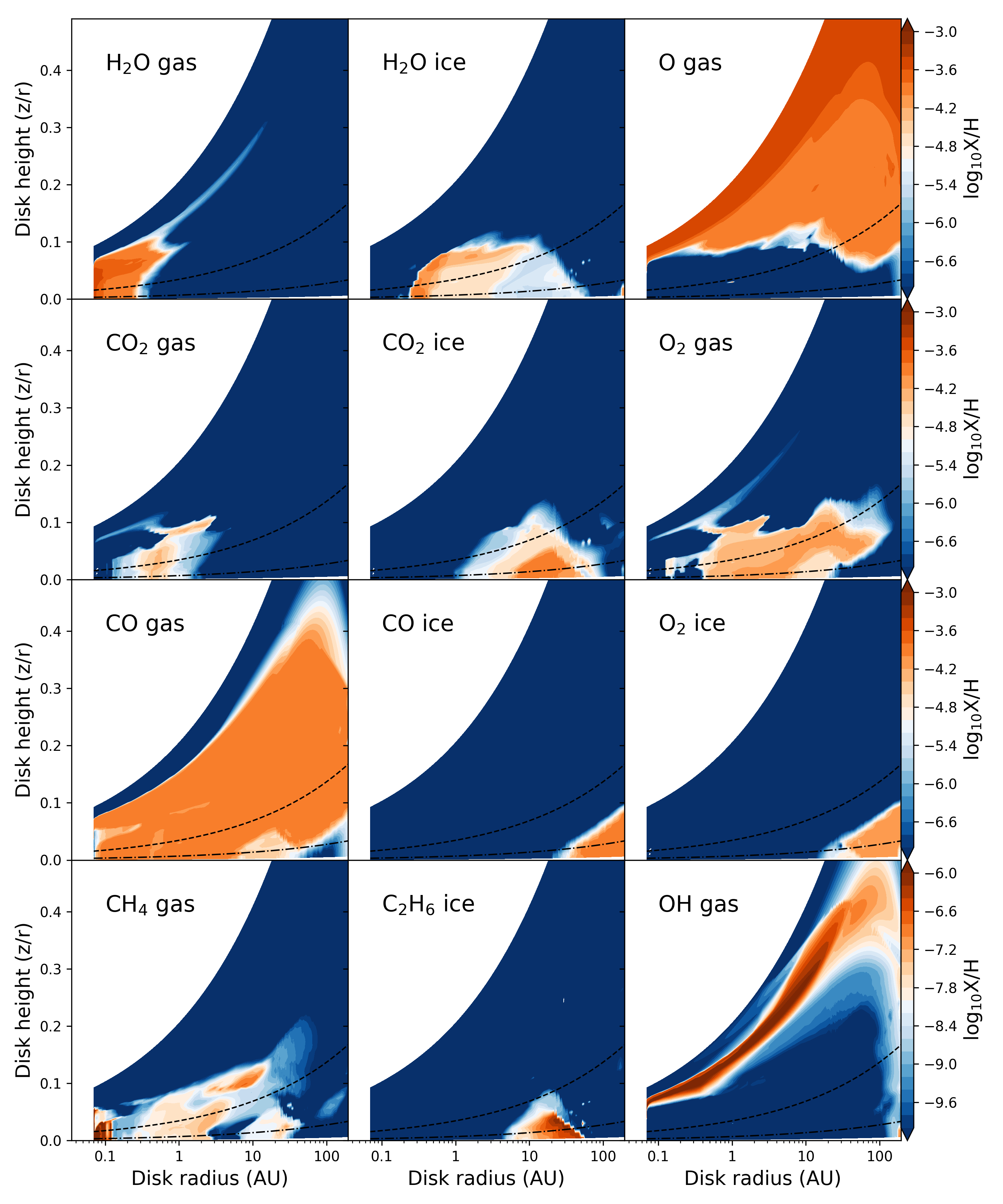}
\caption{Abundance for different relevant species in our chemical model. Note that the bottom colour contour in the bottom row is different than the rest of the figure.}
\label{fig:app}
\end{figure*}

In Figures \ref{fig:app} we note the abundance relative to the number of hydrogen atoms in either log scale or linear (for CO). We highlight in the text the enhancement of CO$_2$ ice seen in the figure caused by the conversion of CO gas into CO$_2$ on the dust grains. We see a corresponding reduction of CO gas outward of 8 AU near the midplane of the disk.

Also in Figure \ref{fig:app} the CO ice surface is easily seen outside of the midplane ice line of $\sim40$ AU. We note that the region of high C/O in Figure \ref{fig:chemmodel02} occurs just above the water ice surface, and corresponds to a warmer region with a slight increase of CH$_4$. Note that the last row of figures are on a different colour scale.

\end{document}